%% file: main.tex
\documentclass[a4paper,usenatbib,times]{mnras}
\usepackage{natbib,graphicx,amsmath,amssymb,times,fleqn,url,mathtools,txfonts,pstricks}

\usepackage[utf8]{inputenc}

\newcommand{\LL}{\mathcal{L}}
\newcommand{\A}{\mathcal{A}}
\newcommand{\B}{\mathcal{B}}

\newcommand{\R}{\mathbb{R}}

\title[Dust settling of small grains]{On the settling of small grains in dusty discs: analysis and formulas}

\author[Laibe et al.]{Guillaume Laibe$^{1}$\thanks{guillaume.laibe@ens-lyon.fr},Charles-Edouard Br\'ehier $^{2}$\thanks{brehier@math.univ-lyon1.fr} and Maxime Lombart$^{1}$\\
$^{1}$\'Ecole normale sup\'erieure de Lyon, CRAL, UMR CNRS 5574, Universit\'e de Lyon , 46 All\'ee d'Italie, 69364 Lyon Cedex 07, France \\
$^{2}$ Univ Lyon, CNRS, Universit\'e Claude Bernard Lyon 1, UMR5208, Institut Camille Jordan, F-69622 Villeurbanne, France
}

\pagerange{\pageref{firstpage}--\pageref{lastpage}} \pubyear{2019}

\begin{document}
\include{journaux}

\label{firstpage}
\bibliographystyle{mn2e}
\maketitle

\begin{abstract}
Instruments achieve sharper and finer observations of micron-in-size dust grains in the top layers of young stellar discs. To provide accurate models, we revisit the theory of dust settling for small grains, when gas stratification, dust inertia and finite correlation times for the turbulence should be handled simultaneously. We start from a balance of forces and derive distributions at steady-state. Asymptotic expansions require caution since limits do not commute. In particular, non-physical bumpy distributions appear when turbulence is purely diffusive. This excludes very short correlation times for real discs, as predicted by numerical simulations.
\end{abstract}

\begin{keywords}
Planets and satellites: formation, Diffusion, Turbulence, Methods: analytical, Methods: numerical.
\end{keywords}

\section{Introduction}
\label{sec:intro}

Details of the structure of dusty discs are now accessible by the mean of instruments such as the Atacama Large (sub)Millimetre Array ALMA (e.g. \citealt{vanderMarel2013,HLTau2015,Andrews2018}), the Spectro-Polarimetric High-contrast Exoplanet REsearch instrument SPHERE/VLT (e.g. \citealt{Benisty2015,Avenhaus2018}) or the Gemini Planet Imager Gemini/GPI (e.g. \citealt{Laws2020}). Spatial differentiation between gas and dust grains is evidenced, both in the midplane of the disc (radial drift) or in the vertical direction (vertical settling). Centimetre-in-size pebbles have attracted lot of attention as they provide primordial material to form planetary cores (e.g. \citealt{Chiang2008,Testi2014}). Small micron-in-size grains are as important (e.g. \citealt{Apai2004,Furlan2006,Dent2013,Espaillat2014,Maaskant2015}), since they are often used as a proxy for the gas. They also set the charge and thermal balances of the disc and radiate polarized light. Hence the need of an accurate description of vertical distributions of small particles.

Primary theories of dust settling (e.g. \citealt{Hoyle1960,Kusaka1970,Cameron1973, Adachi1976,HW1977,Coradini1980}) have emerged with the development of the planetary nebulae hypothesis \citep{Mendoza1966,Safronov1969}. Further developments of the Minimum Mass Solar Nebulae models (e.g. \citealt{CameronPine1973,Weiden1977b,Hayashi1981}) sparked models coupling settling to growth \citep{Weiden1980,Nakagawa1981}. The idea that turbulence sustains dust stirring \citep{Cuzzi1993} emanated from observations of Spectral Energy Distributions of T-Tauri objects \citep{KH1987} concomitant to the rediscovery of the magneto-rotational instability \citep{Balbus1991}.

The seminal theory of dust settling was established by  \citet{Dubrulle1995}. Turbulence is treated by the mean of a Fokker-Planck equation, an approach that resulted in a widely-used model to estimate dust scale heights in discs. Soon after, \citet{Dullemond2004b,DullemondDom2005} pioneered models of dust settling coupled to Monte-Carlo methods for radiative transfer, a technic extended to ray tracing by \citet{Pinte2006,Pinte2007}. \citet{Tanaka2005} modelled spectral energy distributions expected from the interplay between settling and coagulation. In parallel, several aspects of dust settling were quantified with (magneto)-hydrodynamical simulations: the role of dust feed-back \citep{BF2005,Johansen2005,Johansen2006}, turbulence \citep{Takeuchi2002,Carballido2006,Fromang2006,Fromang2009,Ciesla2010,Turner2010,Charnoz2011,Johansen2011,Carballido2011,Zhu2015,Stoll2016,Lin2019}, and grain growth/fragmentation \citep{Zsom2011}. Analytic or semi-analytic models were refined to understand the role played by different drag regime \citep{GaraudLin2004}, refined models of turbulence \citep{SH04,Jacquet2013,Ormel2018}, turbulent dead zones \citep{Ciesla2007}, turbulent correlations \citep{Youdin2007}, grain growth \citep{Laibe2014c} or winds \citep{Riols2018}. These models are widely used to infer the properties of the disc from observations (e.g. \citealt{deBoer2017,Dullemond2018,Sengupta2019,Greenwood2019,Liu2019}).

However, we still lack an analytic formula for the distribution of small dust grains that encompass gas stratification, dust inertia and finite correlation times. To obtain such a recipe, we depart from the historical Fokker-Planck approach and start directly from a balance of forces on a dust grain (Sect.~\ref{sec:physics}). We obtain a system of stochastic differential equations that we analyse, in the spirit of \citet{Ormel2018} (Sect.~\ref{sec:maths}). Results are validated against numerical simulations in Sect.~\ref{sec:numerics} and discussed in Sect.~\ref{sec:discussion}.

\section{Physical model}
\label{sec:physics}

\subsection{Balance of forces}

We consider a non-magnetic non self-gravitating vertically isothermal disc made of gas and dust. We denote  by $r$ and $z$ the radial and the vertical coordinates respectively. The central star is modelled as a point mass $M_{\star}$, such that the vertical component $g_{z}$ of its gravitational field is
\begin{equation}
g_{z} = - \frac{\mathcal{G}M_{\star}z}{\left(r^{2} + z^{2} \right)^{3/2}}.
\label{eq:gz}
\end{equation}
 The gas is supposed to be inviscid and the disc to be at vertical hydrostatic equilibrium. Thus, at a distance $r$ from the central star, the gas density is
\begin{align}
\rho_{\rm g} \left(r,z\right) & = \rho_{\rm g 0} \left(r\right) \mathrm{e}^{\displaystyle - \frac{\mathcal{G} M_{\star}}{c_{\rm s}^{2}\left(r\right)} \int_{0}^{z} g\left( z'\right) \mathrm{d}z' } , \\
& = \rho_{\rm g 0} \left(r\right) \mathrm{e}^{\displaystyle - \frac{\mathcal{G} M_{\star}}{c_{\rm s}^{2}\left(r\right)} \left[\frac{1}{r} - \frac{1}{\sqrt{r^{2}+z^{2}}} \right]} ,
\label{eq:simpcsr}
\end{align}
where $\rho_{\rm g 0}$ and $c_{\rm s}$ denote the gas density in the midplane of the disc and the sound speed at a distance $r$ from the star respectively (e.g. \citealt{Laibe2012}). In the limit $z \ll r$, $g_{z}$ linearises into
\begin{equation}
g_{z} \simeq - \frac{\mathcal{G}M_{\star}z}{r^{3}} = - \Omega^{2} z ,
\label{eq:simplegrav}
\end{equation}
a spring-like force of frequency the orbital frequency of the disc. Under this approximation, Eq.~\ref{eq:simpcsr} reduces to  
\begin{equation}
\rho_{\rm g} \left(r,z\right) = \rho_{\rm g 0} \left(r\right) \mathrm{e}^{\displaystyle - \frac{z^{2}}{2 H^{2}}} ,
\label{eq:strat}
\end{equation}
where $H \equiv \Omega^{-1} c_{\rm s}$ denotes the pressure scale height of the gas. The typical aspect ratio $H/r$ of observed discs is of order $\sim 0.1$. Close to the midplane of the disc ($z \ll H$), Eq.~\ref{eq:strat} becomes
\begin{equation}
\rho_{\rm g} \left(r,z\right) = \rho_{\rm g 0} + {\rm O} \left( z^{2}/H^{2} \right) .
\label{eq:simpledens}
\end{equation}
Dust grains are assumed to be compact, homogeneous and of spherical shape with radius $s$. Grains are uncharged, although this assumption might not be correct anymore for $z \lesssim 3H$ (e.g. \citealt{Bai2009}). The mass of the grain is therefore $m_{\rm d} = \frac{4}{3}   \pi \rho s^{3}$, where $\rho$ denotes the intrinsic density of the grain material -- typically a few g.cm$^{-1}$. In typical classical T-Tauri star discs, the collisional mean free path of the gas is larger than the size of the grain. The drag force $\mathbf{f}_{\rm d}$ exerted by the gas on grains is
\begin{equation}
\mathbf{f}_{\rm d} = - m_{\rm d} \frac{\left(\mathbf{v}_{\rm d} - \mathbf{v}_{\rm g} \right)}{t_{\rm s}} ,
\end{equation}
where $t_{\rm s}$ denotes the drag stopping time, i.e. the typical time for dust grains to reach gas velocity. The stopping time depends on the gas and dust parameters according to
\begin{equation}
t_{\rm s} = \frac{\rho s}{\rho_{\rm g} c_{s}} \sqrt{\frac{\pi \gamma}{8}} ,
\label{eq:ts}
\end{equation}
where $\gamma$ denotes the adiabatic index of the gas \citep{Epstein1924,Baines1965,Whipple1972}. Combining Eqs.~\ref{eq:simpcsr} and \ref{eq:ts},
\begin{equation}
t_{\mathrm{s}} \left( z \right) = t_{\mathrm{s} 0} \, \mathrm{e}^{\displaystyle  \frac{\mathcal{G} M_{\star}}{c_{\rm s}^{2}\left(r\right)} \left[\frac{1}{r} - \frac{1}{\sqrt{r^{2}+z^{2}}} \right]} \simeq t_{\mathrm{s} 0} \, \mathrm{e}^{\displaystyle \frac{z^{2}}{2 H^{2}}} ,
\label{eq:time_stop}
\end{equation}
where $t_{\mathrm{s} 0}$ denotes the stopping time in the midplane. Hence, grains decouple very efficiently in the high atmosphere of the disc where the gas density drops, and the stopping time is an increasing function of the vertical height inversely proportional to gas density.

Gravity from the star and gas drag are the two main relevant forces for this problem. Additional contributions such as radiation forces, magnetic forces or other hydrodynamical forces are negligible \citep{LP12a}. Quadratic corrections for supersonic drag are not expected to play any sensible contribution in this problem and are neglected \citep{Kwok1975}. The ratio of the timescales between the vertical and the radial timescale is of order $\left(H/r \right)^{2} \sim 0.01$, justifying treating $r$ as a constant \citep{laibe2014a}. This assumption holds whenever $z$ is small enough for the conservation of angular momentum to remain valid up to second order in $z/r$. 
 
From the expression of the stopping time given by Eq.~\ref{eq:time_stop}, the balance of forces for single dust grain provides
\begin{equation}
\ddot{z} + \frac{\left( \dot{z} - v_{\mathrm{g}, z} \right)}{ t_{\mathrm{s}, 0}} \mathrm{e}^{\displaystyle - \frac{\mathcal{G} M_{\star}}{c_{\rm s}^{2}\left(r\right)} \left[\frac{1}{r} - \frac{1}{\sqrt{r^{2}+z^{2}}} \right]} + \frac{\mathcal{G}M_{\star}z}{\left(r^{2} + z^{2} \right)^{3/2}} = 0 .
\label{eq:eq_dim}
\end{equation}
We now introduce the dimensionless quantities $Z \equiv z/ H$, $T \equiv t / \Omega^{-1}$ and $\dot{Z} = \dot{z} / c_{\rm s}$. Note that $T$ denotes the time in units of the orbital period and \textit{not} the temperature. We scale also the gas velocity by its sound speed, i.e. $V_{\rm g} = v_{\rm g} / c_{\rm s}$. We denote by the constant $\phi = H/r$ the local aspect ratio of the disc. The Stokes number $S_{\! \rm t} \equiv \Omega t_{s}$ measures the relative contribution between gas drag and gravity. From Eq.~\ref{eq:time_stop}, it increases with vertical height as
\begin{equation}
S_{\! \rm t} = \displaystyle S_{\! \rm t 0} \, \mathrm{e}^{ Z^{2} / 2} ,
\end{equation}
where $S_{\! \rm t 0}$ denotes the Stokes number in the midplane of the disc. We note that grains reach $S_{\! \rm t} = 1$ for $Z = \displaystyle \sqrt{-2 \ln   S_{\! \rm t 0} }$, i.e. a few pressure scale heights even for tiny values of $S_{\! \rm t 0}$. Starting from Eq.~\ref{eq:eq_dim} and rearranging the terms, one obtains the equation of motion for a single grain:
\begin{equation}
\ddot{Z} + \displaystyle S_{\! \rm t 0}^{-1}  f_{\phi}\left(Z \right) \dot{Z} + g_{\phi}\left(Z\right)  = \displaystyle S_{\! \rm t 0}^{-1}   f_{\phi}\left(Z \right) V_{\rm g} ,
\label{eq:motion}
\end{equation}
where
\begin{eqnarray}
f_{\phi}\left(Z \right) & \equiv &  \mathrm{e}^{ - \frac{1}{\phi^{2}} \left[ 1- \left( 1 + \left(\phi Z \right)^{2} \right)^{-1/2} \right]  } , \label{eq:def_f}\\
g_{\phi}\left(Z\right) & \equiv & \frac{Z}{\left(1 + \phi^{2}Z^{2} \right)^{3/2}} . \label{eq:def_g}
\end{eqnarray}
Effects of vertical stratification are still encapsulated in the Taylor expansion of Eq.~\ref{eq:motion} with respect to the small parameter $\phi^{2}\sim 0.01$ 
\begin{equation}
\ddot{Z} + S_{\! \rm t 0}^{-1}  \mathrm{e}^{ - Z^{2} / 2} \dot{Z} + Z = S_{\! \rm t 0}^{-1}  \mathrm{e}^{ - Z^{2} / 2} V_{\rm g} .
\label{eq:eq_dim2}
\end{equation}
The final step of the model consists of modelling the turbulent velocity of the gas $V_{\rm g}$, which appears in the right-hand side of Eq.~\ref{eq:motion}. In the limiting case of a laminar flow, $V_{\rm g} = 0$ and Eq.~\ref{eq:motion} reduces to the well-known equation for vertical settling in laminar discs (e.g. \citealt{Laibe2014c}). 

\subsection{Modelling dusty turbulence}
\label{sec:turb_models}

\subsubsection{Lagrangian turbulence}
\label{sec:lagturb}

The gas velocity is unknown since no exact analytic solution for turbulence in a disc -- and turbulence in general -- are known. However, statistical properties of turbulence can be inferred from laboratory, numerical experiment or theory, and turbulent fluctuations can be modelled using stochastic processes, independently from the origin of the turbulence itself. In a seminal study, \citet{Thomson1987} proved that the only expression of $v_{\rm g}$ that is consistent with Kolmogorov turbulence and the hydrodynamical equations is
\begin{equation}
\frac{\mathrm{d}v_{\rm g}}{\mathrm{d}t} = -\frac{v_{\rm g}}{t_{\rm e}}  + \frac{\sqrt{D}}{t_{\rm e}}\dot{w} ,
\label{eq:turbvg}
\end{equation}
where $t_{\rm e}$ denotes the Lagrangian timescale of the turbulence, $D$ is the turbulent diffusivity (in units m$^{2}$s$^{-1}$). $w$ is a Wiener process, such that its derivative is a white noise such that
\begin{eqnarray}
\left< \dot{w} (t) \right> & = &  0, \label{eq:Wiener1} \\
\left< \dot{w} (t) \, \dot{w} (t') \right> & = &  \delta(t - t') , \label{eq:Wiener2}
\end{eqnarray}
where $\delta$ denotes the Dirac distribution and the notation $\left< \cdot \right>$ is the expectation operator (see also \citealt{sawford1984,WilsonSawford}). Eq.~\ref{eq:turbvg} describes turbulent fluctuations from a Lagrangian point of view \citep{Taylor1921}. From Eq.~\ref{eq:turbvg}, the gas velocity can be rewritten
\begin{equation}
v_{\rm g} = \zeta\left(t,t_{\rm e},D \right) ,
\label{eq:vg_OU}
\end{equation}
where $\zeta$ is a stationary Ornstein-Uhlenbeck process defined by
\begin{eqnarray}
\left< \zeta (t,t_{\rm e},D) \right> & = & 0, \label{eq:OU1} \\
\left< \zeta (t,t_{\rm e},D) \, \zeta (t',t_{\rm e},D) \right> & = &  \frac{D}{2 t_{\rm e}} \rm{e}^{-\frac{\vert t - t' \vert}{t_{\rm e}}} . \label{eq:OU2}
\end{eqnarray}
Eq.~\ref{eq:turbvg} defines a model of turbulence with two parameters, $D$ and $t_{e}$. In discs, $t_{\rm e}$ is typically of order one orbital period, since turbulent vortices are stretched out by differential rotation in a few orbits (e.g. \citealt{Beckwith2011}). From Eq.~\ref{eq:OU2}, $D$ is related to the auto-correlation of the turbulent noise according to
\begin{equation}
D = 2 \int_{0}^{+\infty} \left< v_{\rm g}\left( 0 \right) v_{\rm g}\left( t \right)\right> \mathrm{d}t .
\label{eq:defD}
\end{equation}
Eq.~\ref{eq:defD} can alternatively be seen as a definition of the turbulent diffusivity, useful in practice to measure $D$ in numerical simulations. The Wiener-Khinchin theorem ensures that the power spectrum of the turbulent velocity field $S \! \left( \omega \right)$ is the Fourier transform of this autocorrelation function, i.e.
\begin{equation}
S \left( \omega \right)  =  \frac{1}{2 \pi} \int_{-\infty}^{+\infty} \mathrm{e}^{- i \omega t} \left< v_{\rm g}\left( 0 \right) v_{\rm g}\left( t \right)\right> \mathrm{d}t =  \frac{D}{2\pi \left( 1 + \omega^2 t_{\rm e}^2 \right)} .
\label{eq:WK}
\end{equation}
Thus, in the inertial subrange ($ \omega^2 t_{\rm e}^2 \gg 1$), we have $S(\omega) \propto \omega^{-2} $, whose equivalent in the wavelength space is $\tilde{S}(k)\propto k^{-5/3}$ \citep{Batchelor1950}. From Eq.~\ref{eq:WK}, the standard deviation of the velocity fluctuation $\sigma$ is
\begin{equation}
\sigma^{2} \equiv \int_{-\infty}^{+\infty} S\left( \omega \right) \mathrm{d} \omega = \frac{D}{2t_{\rm e}} .
\label{eq:FD}
\end{equation}
Physically, Eq.~\ref{eq:FD} is a turbulent fluctuation-dissipation theorem. 

In astrophysics, the turbulent activity of a disc is often parametrised by a constant denoted by $\alpha$ \citep{ShakuraSunyaev1973}. In this work, we define $\alpha$ according to
\begin{equation}
\alpha \equiv \frac{D}{2 c_{\rm s} H} ,
\label{eq:def_alpha}
\end{equation}
to be consistent with previous studies on dust diffusivity (e.g. \citealt{Fromang2006}). Combining Eqs.~\ref{eq:OU2} and \ref{eq:def_alpha}, Eq.~\ref{eq:vg_OU} can be rewritten as
\begin{equation}
v_{\rm g} / c_{\rm s} = \sqrt{2 \alpha} \, \zeta\left(T, \tau_{\rm e}, 1 \right) ,
\end{equation}
where $\tau_{\rm e} \equiv t_{e} \Omega$. In the literature, the \textit{same notation} $\alpha$ has been used to denote \textit{different} dimensionless physical quantities, all related to the turbulent activity of the disc and being therefore of the same order of magnitude. The parameter $\alpha$ may be used e.g. for quantities measuring the efficiency of the transport of angular momentum, the intensity of the velocity fluctuations or the turbulent diffusivity (e.g. \citealt{Arena2013}). For a quantitative use of our results, values of $\alpha$ should either be directly measured using Eq.~\ref{eq:defD} or deduced from an alternative measurement of the turbulent activity of the disc and a coefficient of proportionality which has been calibrated independently. The ratio between turbulent and thermal pressure is of order $\sqrt{\alpha}$. Hence, Eq.~\ref{eq:simpcsr} remains a valid expression for the density profile of the disc.

\begin{figure}
  \includegraphics[width=.45 \textwidth]{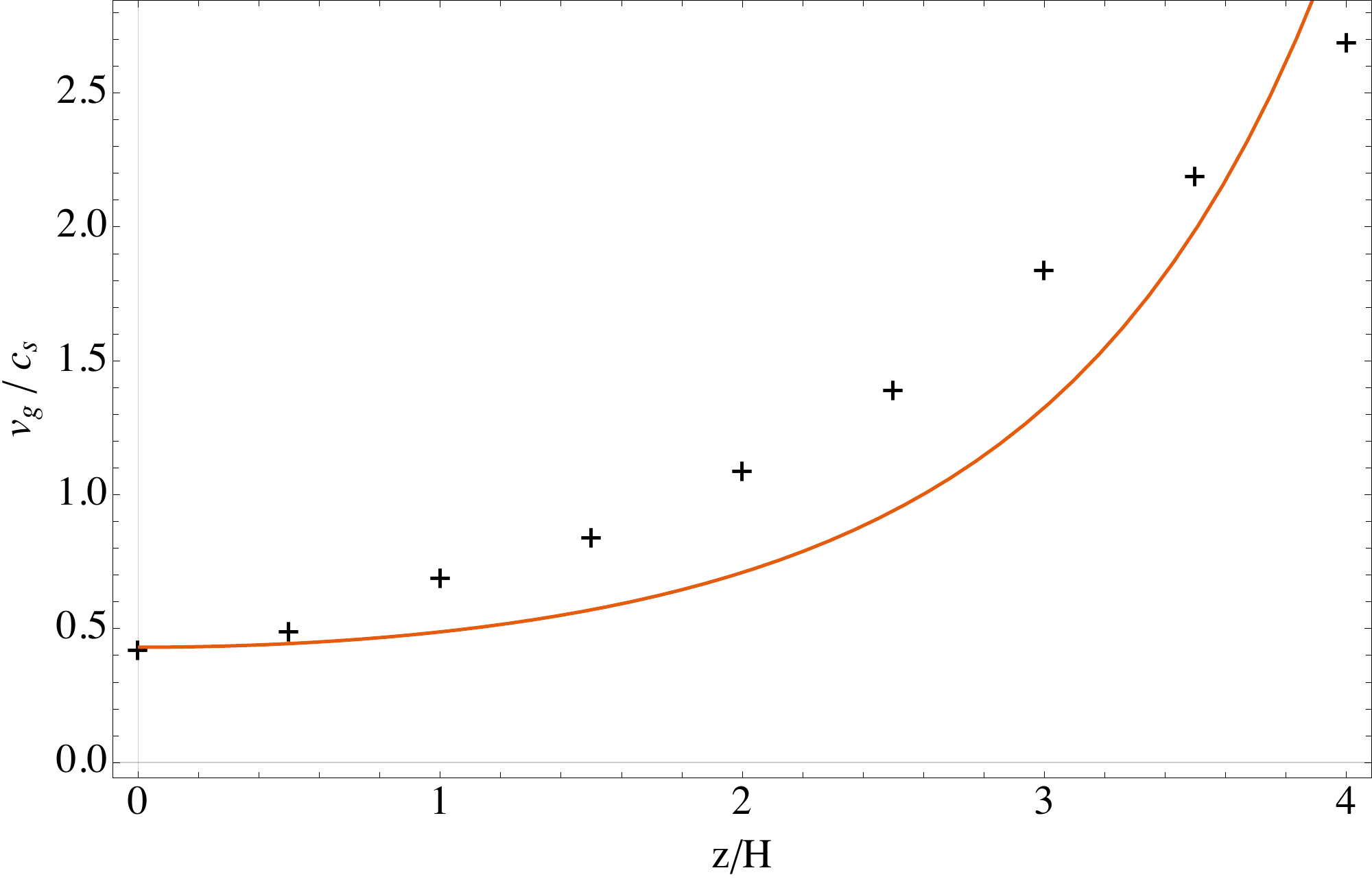}
   \caption{Comparison between the toy model $\alpha(z) = \alpha_{0}  / \sqrt{\rho_{\rm g}\left(z \right) / \rho_{0}}$ and values of $\alpha$ measured directly from the MHD numerical simulations of \citet{Fromang2009}. The global trend is reproduced with maximum errors reaching $\simeq 50\%$, which is sufficient for this study.}
   \label{fig:alphaz}
\end{figure}
The vertical dependency of $\alpha$ can be inferred from numerical simulations of magneto-hydrodynamical turbulence (e.g. \citealt{Miller2000,Fromang2009,Fromang2010}). For numerical tractability, simulations are performed in a local shearing-box that extends vertically over a few pressure scale heights. Fig.~\ref{fig:alphaz} displays values of $\alpha$ measured by \citet{Fromang2009}. A generic feature is that $\alpha$ increases with $z$. No first-principle model exists so far to prescribe $\alpha(z)$. Alternative recipes have been proposed to mimic this behaviour  (e.g. \citealt{Ciesla2010,Ormel2018}).  In this study, we use for convenience and tractability  a very crude but parameter-free parametrisation of $\alpha \left(z \right)$
\begin{equation}
\alpha  = \alpha_{0} \left(\rho_{0} / \rho \right)^{1/2} .
\label{eq:alphaz_full}
\end{equation}
This assumption ensures that the density of turbulent energy $\rho_{\rm g} v_{\rm g}^{2}$ remains finite and roughly constant in a vertical slab of the disc. The model is therefore compatible with a steady-state, since no further turbulent processes are required to smooth out local energy gradients. The agreement between Eq.~\ref{eq:alphaz_full} and numerical simulations is quite reasonable (Fig.~\ref{fig:alphaz}). Although errors may reach $\simeq 50 \%$, the model is conservative for our study since it  enhances slightly the eventual role played by a positive value of the vertical gradient of $\alpha$. The prescription may probably be incorrect for $z\gtrsim 3H$ \citep{Fromang2009}. This does not affect significantly our results since we find almost no grain at these heights. In dimensionless quantities, we denote
\begin{equation}
\alpha = \alpha_{0} h^{2}\left(Z\right) .
\label{eq:alphaz}
\end{equation}
A value of $h = f^{-1/4}$ corresponds to Eq.~\ref{eq:alphaz_full}. 

Turbulence could have alternatively been described by the turbulent velocity $\sigma_{\rm \!t}$ and the mean rate of dissipation of turbulence kinetic energy $\epsilon_{\rm t}$, where $t_{\rm e} = \frac{2 \sigma_{\rm \!t}^{2}}{C_{0} \epsilon_{\rm t}} $, $ D = \frac{\left( 2 \sigma_{\rm \!t}^{2} \right)^{2}  }{C_{0} \epsilon_{\rm t}}$, and $C_{0}$ is a constant to be calibrated \citep{Thomson1987}. Thus, $\sigma_{\rm \!t} = {\rm O}\left( \sqrt{\alpha} c_{\rm s} \right)$ and $\epsilon_{\rm t} = {\rm O}\left( \alpha c_{\rm s}^{3} / H \right)$. The turbulent viscosity $\nu$ scales like $\nu = {\rm O}\left(\sigma_{\rm \!t}^{2}/\epsilon_{\rm t} \right) = {\rm O}\left(\alpha c_{\rm s} H\right)$, consistently with a Sakura and Sunyaev prescription, for which $\nu = \alpha_{\rm SS} c_{\rm s} H$. It is found in numerical simulations that in protoplanetary discs, $\alpha \simeq \alpha_{\rm SS} \simeq 10^{-4} - 10^{-2}$. Further refined stochastic models including multiple turbulent timescales have been used in the context of aerosols and suspensions (e.g. \citealt{Shao1995,Pope2002}). Effects of anisotropy may also be described by other turbulent parameters (e.g. \citealt{BP1999,Ogilvie2001,Lodato2008,Balbus2011}). These refinements are not expected to have any significant impact in our study.

\subsubsection{Equations of motion}
\label{sec:SDE}

Combining Eqs.~\ref{eq:motion} -- \ref{eq:def_g}, Eq.~\ref{eq:turbvg} and Eq.~\ref{eq:alphaz} in a dimensionless form, one obtains the system
\begin{eqnarray}
\mathrm{d}Z & = & V \mathrm{d}T  , \label{eq:sto1} \\
\mathrm{d}V + S_{\! \rm t 0}^{-1}  f_{\phi}(Z) V \mathrm{d}T + g_{\phi}(Z) \mathrm{d}T  & = & S_{\! \rm t 0}^{-1} f_{\phi}(Z) h(Z) \sqrt{2\alpha_{0}}  \,  \xi\mathrm{d}T \label{eq:sto2} \\
\mathrm{d}\xi & = & -\frac{\xi}{\tau_{\rm e}}dT + \frac{\mathrm{d} w}{\tau_{\rm e}}  . \label{eq:sto3}
\end{eqnarray}
The system of equations Eqs.~\ref{eq:sto1} -- \ref{eq:sto3} is stochastic and the dust scale height in steady-state is subsequently defined in a probabilistic way as the variance of the dust distribution at large times, i.e.
\begin{equation}
H_{\rm d}  \equiv \sqrt{\left<ZZ \right>}_{T=+\infty} .
\label{eq:defhd}
\end{equation}
Gravity, which confines dust particle close to the midplane, is encompassed in the function $g_{\phi}$. in In a real disc, this confinement is weaker than if it were operated by the osculating harmonic potential of the midplane. On the one hand, gas drag dissipates the mean kinetic energy of the grain through the second term of the left-hand side of Eq.~\ref{eq:sto2}. This makes grains settle to the midplane, which is the bottom of the potential well. On the other hand, gas drag couples the grain to the stirring turbulent fluctuations of the gas through the driving term of the right-hand side of Eq.~\ref{eq:sto2}. 

Density stratification of the gas is encoded in the function $f_{\phi}$. From Eqs.~\ref{eq:sto1}-- \ref{eq:sto3}, stratification affects the dynamics of the grain in two ways. Firstly, grains having small Stokes numbers $S_{\! \rm t 0}\ll 1$ in the midplane may have Stokes numbers $S_{\! \rm t} = S_{\! \rm t 0}/f_{\phi}$ larger than unity in the top-layers of the disc. Thus, the dynamics of those grains may counter-intuitively be gravity-dominated. Secondly, grains couple and react more efficiently to turbulent stirring close to the midplane of the disc. If the gradient of the product $f_{\phi} h$ is negative, grains receive stronger turbulent kicks from the bottom of the disc than from the top layers. For smooth vertical profiles of $\alpha$, this differential effect is the strongest close to the inflection point of the density profile, i.e. one pressure scale height for the Gaussian profile. Hence, stratification affects the stirring of small grains ($S_{\! \rm t 0} \sim \alpha \ll1$) and can not be neglected. Its effects are the strongest in the top layers of the disc, where the dynamics is gravity-dominated and submitted to a large differential driving. 

\subsection{Link with previous works}
\label{sec:compendium}
\subsubsection{Strong drag approximation}

A first approximation for Eqs.~\ref{eq:sto1} -- \ref{eq:sto3} consists of assuming that grains are small enough for the dynamics to be always drag dominated and that $\ddot{Z} \ll S_{\! \rm t 0}  f_{\phi}\left(Z \right) \dot{Z}$ in Eq.~\ref{eq:motion}. For the sake of clarity, we shall now use the approximations of Eqs.~\ref{eq:simplegrav}--\ref{eq:strat} for $f_{\phi}$ and $g_{\phi}$ and a constant viscosity ($h = 1$) to illustrate the effect of this approximation since it does not affect the nature of our conclusions. The evolution of dust grains is therefore governed by the equation 
\begin{equation}
\mathrm{d} Z = - S_{\! \rm t 0} Z \mathrm{e}^{Z^{2}/2} \mathrm{d}T + \sqrt{2 \alpha} \mathrm{d} \xi .
\label{eq:dragdom}
\end{equation}
 For a purely diffusive process $\xi$, Eq.~\ref{eq:dragdom} is equivalent to the following Fokker-Planck equation (e.g. \citealt{Risken1989fpe})
\begin{equation}
\frac{\partial p}{\partial T} = \frac{\partial}{\partial Z} \left( S_{\! \rm t 0} Z \mathrm{e}^{Z^{2}/2} p \right) + \alpha \frac{\partial^{2}p}{\partial Z^{2}}.
\label{eq:fpe_simp}
\end{equation}
In the Fokker-Planck formalism, a definition of the dust scale height equivalent to Eq.~\ref{eq:defhd} is %
\begin{equation}
H_{\rm d}  = \left( \int_{-\infty}^{+\infty} \!\!\! \int_{-\infty}^{+\infty}p(+\infty,Z,V) Z^{2} \, \mathrm{d}V \mathrm{d}Z \right)^{1/2} .
\label{eq:defhd_fp}
\end{equation}
Eq.~\ref{eq:fpe_simp} does not depend on the velocity $V$ anymore, and its steady-state solution is (e.g. \citealt{NLoscillBook,Fromang2009})
\begin{equation}
p(z) \propto \mathrm{e}^{\displaystyle \int_{0}^{z} - \frac{Z' \mathrm{e}^{Z'^{2}/2}}{\alpha / S_{\! \rm t 0}} \mathrm{d}Z' } = \mathrm{e}^{ \displaystyle  -\frac{ \mathrm{e}^{Z^{2}/2} }{ \alpha / S_{\! \rm t 0} }} ,
\label{eq:sol_fpesimp}
\end{equation}
which gives the dust scale height on an integral form
\begin{equation}
H_{\rm d}  = \left( \int_{-\infty}^{+\infty} \!\!\! Z^2 \mathrm{e}^{ \displaystyle  -\frac{ \mathrm{e}^{Z^{2}/2} }{ \alpha / S_{\! \rm t 0} }} \mathrm{d} Z \,\,\,/ \int_{-\infty}^{+\infty} \!\!\! \mathrm{e}^{ \displaystyle  -\frac{ \mathrm{e}^{Z^{2}/2} }{ \alpha / S_{\! \rm t 0} }} \mathrm{d} Z \right)^{1/2}
\end{equation}

The parameter $\alpha / S_{\! \rm t 0} $ appears naturally as the relevant quantity to measure whether dust grains are significantly sensitive to the turbulent activity of the gas or not.
\\
In the dust distribution given by Eq.~\ref{eq:sol_fpesimp}, small grains remain confined within almost three pressure scale heights around the midplane. Indeed, the low gas density in the top layers of the disc reduces drastically the efficiency of turbulent driving, preventing the particles to escape. Eq.~\ref{eq:dragdom} shows that gas stratification acts as a stiff effective potential $\mathcal{V}_{\rm eff}(Z)\equiv S_{\! \rm t 0} \mathrm{e}^{z^{2}/2}$ that confines the particles close to the midplane. As expected, the distribution Eq.~\ref{eq:sol_fpesimp} corresponds to the Boltzmann distribution
\begin{equation}
p(z) \propto \mathrm{e}^{\displaystyle -\mathcal{V}_{\rm eff}(Z) / \alpha} .
\label{eq:boltz}
\end{equation}
In Eq.~\ref{eq:boltz}, $\alpha$ is the dimensionless form of the turbulent energy $ \alpha c_{\rm s}^{2} $. It is not possible to obtain a closed-form expression for the dust scale height in stationary regime from Eq.~\ref{eq:sol_fpesimp}. When the particles are close enough to the midplane of the disc, i.e. when $\alpha / S_{\! \rm t 0} \ll 1$, Eqs.~\ref{eq:dragdom} and \ref{eq:sol_fpesimp} can be linearised, giving
\begin{equation}
p(+\infty,Z) = \sqrt{ \frac{ S_{\! \rm t 0}}{2 \pi \alpha} }  \mathrm{e}^{ \displaystyle -\frac{Z^2}{ 2 \alpha / S_{\! \rm t 0}}} .
\end{equation} 
From Eq.~\ref{eq:defhd}, and integrating over $Z$ only in this case, the analytic expression of $H_{\rm d}$ is
\begin{equation}
H_{\rm d}  =  \sqrt{\alpha / S_{\! \rm t 0}} \, .
\label{eq:Hd_Dubrulle}
\end{equation}
Eq.~\ref{eq:Hd_Dubrulle} is the analytic estimate obtained by \citet{Dubrulle1995} for the dust scale height of particles close to the midplane. Physically, $H_{\rm d}$ is large when turbulence is intense and grains are small, since strong coupling with the gas ensure continuous stirring by the turbulent kicks. Importantly, for $S_{\! \rm t 0} = \alpha$, dust reaches the pressure scale height of the gas ($H_{\rm d} = 1$). This corresponds to Stokes numbers of order $10^{2} - 10^{3}$ in typical discs. Some numerical codes use the expression given by Eq.~\ref{eq:Hd_Dubrulle} since it is easily tractable. However, to overcome the divergence of $H_{\rm d}$ at large coupling parameters, cut-offs for large dust thicknesses need to be enforced, such as 
\begin{equation}
\tilde{H}_{\rm d} \equiv \min \left( \sqrt{\frac{\alpha}{ S_{\! \rm t 0}}}, 1 \right) ,
\label{eq:hd_cut}
\end{equation}
or the smoother variant (e.g. \citealt{Riols2018})
\begin{equation}
\hat{H}_{\rm d} \equiv \left( 1 + \frac{ S_{\! \rm t 0}}{\alpha} \right)^{-1/2} .
\label{eq:hd_cut}
\end{equation}
Although convenient, this approach brings the drawback of not reproducing the step-function aspect of the dust distribution for small grains predicted by Eq.~\ref{eq:sol_fpesimp}.

\subsubsection{Linearisation}

Eqs.~\ref{eq:sto1} -- \ref{eq:sto2} have alternatively been studied by  linearising the function $f_{\phi}$ and $g_{\phi}$ in the limit $Z \ll 1$ according to 
\begin{eqnarray}
\mathrm{d}Z & = & Z \mathrm{d}T  , \label{eq:stolin1} \\
\mathrm{d}V + S_{\! \rm t 0}^{-1} V \mathrm{d}T + Z \mathrm{d}T  & = & S_{\! \rm t 0}^{-1} \sqrt{2\alpha} \, \mathrm{d}\xi \label{eq:stolin2} .
\end{eqnarray}
 This approximation is valid when the dust evolution occurs close to the disc's midplane. \citet{Carballido2006} model turbulence by a white noise, i.e. $\mathrm{d}\xi$ is the Wiener process given by Eqs.~\ref{eq:Wiener1} -- \ref{eq:Wiener2}. Eqs.~\ref{eq:stolin1}--\ref{eq:stolin2} are equivalent to the Fokker-Planck equation
\begin{equation}
\frac{\partial p}{\partial T} + V \frac{\partial p}{\partial Z} + \frac{\partial}{\partial V}\left(\left[ - S_{\! \rm t 0}^{-1} V - Z \right] p \right) - \alpha S_{\! \rm t 0}^{-2}  \frac{\partial^{2} p}{\partial V^{2}} = 0.
\label{eq:lin_FPE}
\end{equation}
A rescaling of Eq.~\ref{eq:lin_FPE} by $\hat{Z} = Z / \sqrt{\alpha / S_{\! \rm t 0}}$ shows that its solution depends only on the product $\alpha / S_{\! \rm t 0}$. The probability density function of the grains converges to the Gaussian distribution
\begin{equation}
p(+\infty,Z,V) = \frac{S_{\! \rm t 0}}{2 \pi \alpha } \mathrm{e}^{-\frac{1}{ \alpha / S_{\! \rm t 0}} \left\lbrace \frac{Z^2}{2} + \frac{V^2}{2} \right\rbrace} ,
\end{equation}
after a typical time $T_{\rm sett} = S_{\! \rm t 0} + S_{\! \rm t 0}^{-1}$ that is the typical settling time in a laminar disc. From Eq.~\ref{eq:defhd}, the dust scale height at equilibrium is 
\begin{equation}
H_{\rm d}  =  \sqrt{\alpha / S_{\! \rm t 0}} .
\label{eq:Dubrulle}
\end{equation}
Remarkably, Eq.~\ref{eq:Dubrulle} provides the same expression than the one obtained in the strong drag approximation (Eq.~\ref{eq:Hd_Dubrulle}). \citet{Youdin2007} have generalised Eq.~\ref{eq:Dubrulle} by including temporal correlations in the model of turbulence. In this case, $\mathrm{d}\xi$ is the Ornstein-Uhlenbeck process given by Eqs.~\ref{eq:OU1} - \ref{eq:OU2}, and the dust scale height depends additionally on the correlation time $\tau_{\rm e}$ according to \citep{Masoliver1993,WangMasoliver1996}
\begin{equation}
H_{\rm d} = \sqrt{\alpha / S_{\! \rm t 0}} \sqrt{\frac{1 +  \tau_{\rm e}/S_{\! \rm t 0}}{  1 + \tau_{\rm e}/S_{\! \rm t 0} + \tau_{\rm e}^{2}  }} .
\label{eq:YL}
\end{equation}
In the limit $\tau_{\rm e} = 0$, Eq.~\ref{eq:YL} reduces to Eq.~\ref{eq:Dubrulle}. For a typical $\tau_{\rm e} = 1$, $H_{\rm d} = \sqrt{\alpha / S_{\! \rm t 0}}\sqrt{\frac{S_{\! \rm t 0} + 1}{2S_{\! \rm t 0} + 1}}$. Hence, in real discs, the qualitative discrepancy between Eq.~\ref{eq:Dubrulle} and Eq.~\ref{eq:YL} is not significant. Moreover, the two models are rigorously equivalent in the small grains limit. Indeed, $H_{\rm d} \simeq \sqrt{\alpha / S_{\! \rm t 0}}$ with an approximation better than one per cent for $S_{\! \rm t 0} < 0.02$. In the limit $\tau_{\rm e} \to +\infty$, the disc is laminar and $H_{\rm d} \to 0$. This case is not relevant in practice.

Importantly, linearised models predict dust scale heights larger than the pressure scale height of the gas for $S_{\! \rm t 0} \lesssim \alpha$ even if physically, there is almost no gas in these layers and thus, almost no turbulent driving. To understand this feature, let us examine closely how equations including stratification behaves against linearisation (i.e. Eqs.~\ref{eq:motion} and \ref{eq:eq_dim2}). The limit of a ``spring-like'' restoring force is obtained by letting the parameter $\phi$ go to zero. This corresponds to the thin cold disc limit. However, the limit of constant damping (Eq.~\ref{eq:simpledens}) can \textit{not} be obtained as an asymptotic behaviour of the equations of evolution with respect to any continuous parameter. Hence, the linearised system of equations models dust particles embedded in an infinite homogeneous vertical slab of gas, whose density is the one of the midplane. Small grains are therefore always scattered efficiently by turbulence wherever their location in the disc, explaining why they are ultimately reaching infinitely high regions. When including stratification, small particles decouple from the gas when they reach a sufficient height and fall back into the minimum of gravitational potential located in the midplane of the disc.

\subsubsection{Diffusion equations}

In \citet{Dubrulle1995}, the dust density is obtained from the Fokker-Planck equation
\begin{equation}
\frac{\partial \rho_{\rm d}}{\partial t} =  \frac{\partial}{\partial z}\left[ z \Omega^{2} t_{\rm s}(z) \rho_{\rm d} \right] + \frac{\partial}{\partial z} \left[ \rho_{\rm g}(z) \kappa_{\rm T}(z) \frac{\partial}{\partial z} \left( \frac{\rho_{\rm d}}{\rho_{\rm g}(z)} \right) \right] ,
\label{eq:FPE_dub}
\end{equation}
where $\kappa_{\rm T}(z)$ is an effective half-diffusivity (to be consistent with Eq.~\ref{eq:def_alpha}). For particles with small Stokes numbers, $ \kappa_{\rm T}(z) = \alpha c_{\rm s} H$  and $ \kappa_{\rm T}(z) \propto z^{-1/2}$ for particles with large Stokes numbers (see \citet{Riols2018} for a detailed discussion of the origin of this equation). In essence, the \citet{Dubrulle1995} model is build on the strong drag approximation and Eq.~\ref{eq:FPE_dub} is equivalent to our Eq.~\ref{eq:fpe_simp} in dimensionless quantities. The discrepancies between the two models can be understood the following way:
\begin{enumerate}
\item The diffusion operator of Eq.~\ref{eq:FPE_dub}, originally introduced by \citet{Morfill1984}, acts on the quantity $\rho_{\rm d}/\rho_{\rm g}$ and not on the quantity $\rho_{\rm d}$. For independent and non-interacting particles, diffusion fluxes smooths gradients of chemical potentials that are proportional to densities. The extrapolation to concentrations is valid only for homogeneous solvent/gas densities. It looks therefore that Eqs.~\ref{eq:sto1}--\ref{eq:sto3} rely on more robust physical bases. We also note that Eq.~\ref{eq:FPE_dub} can not be derived from a balance of forces with stochastic driving. However, the difference between the two equations is only minor, since Eq.~\ref{eq:FPE_dub} writes with our notations
\begin{equation}
\frac{\partial p}{\partial T} = \frac{\partial}{\partial Z} \left( Z\left[  S_{\! \rm t 0}  \mathrm{e}^{Z^{2}/2} + \alpha\right] p \right) + \alpha \frac{\partial^{2}p}{\partial Z^{2}}.
\label{eq:fpe_simp_alt}
\end{equation}
We note the appearance of an extra drift-term for the grains which does not depend on $S_{\! \rm t 0}$. This terms can not be of physical origin, since dust coupled to gas only through gas drag. Anyhow, the extra stir provided by this additional term would affect only tiny grains close to the mid-plane, which are lifted up by turbulence anyway. 

\item The variable diffusivity $\kappa_{\rm T}(z)$ is inherited from an \textit{ad-hoc} concept of eddy classes invoked originally in \citet{Voelk1980}, seven years \textit{before} the work of Thomson. \citet{Riols2018} provide a interpretation for the origin of this term through Reynolds averaging of the dust/gas equations of motion. On the other hand, \citet{Thomson1987} demonstrated that a rigorous way to account eddies of different lifetimes in a Lagrangian descriptions of turbulence is to introduce a finite correlation time $t_{\rm e}$. Eq.~\ref{eq:WK} ensures that the correct spectrum of lifetimes for the turbulent structures in reproduced. In the strong drag approximation, the generalisation of Eq.~\ref{eq:fpe_simp_alt} for finite turbulent times is
\begin{equation}
\frac{\partial p }{\partial T} =  \frac{\partial}{\partial Z} \left( S_{\! \rm t 0} Z \mathrm{e}^{Z^{2}/2} \right) + \sqrt{\alpha / 2} \frac{\partial^{2}}{\partial Z^{2}} \left[  H\left(T,Z \right) p    \right] ,
\label{eq:HM1}
\end{equation}
where $\tilde{H} (Z) = H\left(+\infty, Z \right)$ satisfies
\begin{equation}
\tilde{H} + \tau_{\rm e}\left( \tilde{H} \frac{\partial S_{\! \rm t 0} Z \mathrm{e}^{Z^{2}/2}}{\partial Z} - \frac{\partial \tilde{H} }{\partial Z} S_{\! \rm t 0} Z \mathrm{e}^{Z^{2}/2}  \right) = \sqrt{2 \alpha} .
\label{eq:HM2}
\end{equation}
The general expression for $H\left(Z,t \right)$ is given in \citet{HM1983}. Eqs.~\ref{eq:HM1} -- \ref{eq:HM2} reduce to Eq.~\ref{eq:fpe_simp} when $\tau_{\rm e} = 0$. Interestingly, Eqs.~\ref{eq:HM1} -- \ref{eq:HM2} reduce to 
\begin{equation}
\frac{\partial p }{\partial T} = - \frac{ \partial }{\partial Z}\left(S_{\! \rm t 0} Z p \right) + \frac{\alpha}{1 + S_{\! \rm t 0}\tau_{\rm e}} \left( 1- \mathrm{e}^{-\left[\tau_{\rm e}^{-1} + S_{\! \rm t 0} \right] T} \right)\frac{\partial ^{2} p}{\partial Z^{2}}  ,
\label{eq:HMlin}
\end{equation}
when the equations of evolution are linearised. The equivalent dimensionless diffusivity is $2 \alpha / \left( 1 + S_{\! \rm t 0} \tau_{\rm e} \right)$ and does not depend on $Z$. In the limit $T \to + \infty$, the dust scale height obtained from Eq.~\ref{eq:HMlin} is
\begin{equation}
H_{\rm d} = \sqrt{\frac{\alpha / S_{\! \rm t 0}}{1 + S_{\! \rm t 0}\tau_{\rm e} }} .
\end{equation}
\end{enumerate}

\subsubsection{Conclusion}

So far, no analytic model predicts steady distributions of small grains that can become gravity-dominated in the top layers in stratified discs, where turbulence develops on finite correlation times. Those effects have however been shown to play an important role in structuring the dust layers and are expected to be the most important for the smallest grains. Obtaining a formula which integrates these effects altogether is the goal of the following derivation.

\section{Mathematical analysis}
\label{sec:maths}

\subsection{Rescaling}
\label{sec:rescaling}

For the mathematical analysis, introduce the parameters
\begin{equation}
\epsilon=S_{\! \rm t 0}~,\quad \delta=\sqrt{\tau_{\rm e}S_{\! \rm t 0}}~,\quad \sigma=\sqrt{\alpha_0/S_{\! \rm t 0}}.
\end{equation}
For convenience, we also introduce the parameter $\lambda$ such as
\begin{equation}
\lambda = \delta / \epsilon = \sqrt{\tau_{\rm e} / S_{\! \rm t 0}} .
\end{equation}
Asymptotic analysis is performed in the regime
\begin{equation}
\epsilon\ll 1~,\quad \delta\ll 1,\quad \sigma\sim 1,
\end{equation}
{\it i.e.} the parameters $\epsilon$ and $\delta$ go to $0$ whereas $\sigma$ remains of order $1$. As will be clear below, depending on whether $\delta\ll \epsilon$, $\epsilon\ll \delta$, or $\epsilon \sim \delta$, the limiting equations for $Z$ will be different.

The physical parameters are recovered in terms of the mathematical ones as follows:
\begin{equation}
S_{\! \rm t 0}=\epsilon~,\quad \tau_{\rm e}=\frac{\delta^2}{\epsilon}~,\quad \alpha_0=\sigma^2\epsilon.
\end{equation}

Define $Z'(T)=Z(\epsilon^{-1}T)$, $V'(T)=V(\epsilon^{-1}T)$, $\xi'(T)=\xi(\epsilon^{-1}T)$ and $\zeta'(T)=\delta\epsilon^{-\frac12} \xi'(T)$.

Note that the Stochastic Differential Equation for $\xi'$ is written as
\begin{equation}
d\xi'=-\frac{\xi'}{\tau_{\rm e}\epsilon}\mathrm{d}T+\frac{\mathrm{d}w}{\sqrt{\epsilon}\tau_{\rm e}},
\end{equation}
since in distribution $\bigl(w(\epsilon^{-1}T)\bigr)_{T\ge 0}=\bigl(\epsilon^{-\frac12}w(T)\bigr)_{T\ge 0}$.

As a consequence, the Stochastic Differential for $\zeta'=\delta\epsilon^{-\frac12}\xi'$ is written as
\begin{equation}
d\zeta'=-\frac{\zeta'}{\tau_{\rm e}\epsilon}\mathrm{d}T+\frac{\delta\mathrm{d}w}{\epsilon\tau_{\rm e}}=-\frac{\zeta'}{\delta^2}\mathrm{d}T+\frac{\mathrm{d}w}{\delta}.
\end{equation}
Writing $\xi'=\delta^{-1}\epsilon^{\frac12}\zeta'$ and using the relations between the parameters, one obtains the system (where the notation $f=f_{\phi}$ is used)
\begin{equation}\label{eq:SDE_math}
\begin{cases}
\mathrm{d}Z^{\epsilon,\delta}=\frac{V^{\epsilon,\delta}}{\epsilon}\mathrm{d}t\\
\mathrm{d}V^{\epsilon,\delta}+\frac{f(Z^{\epsilon,\delta})V^{\epsilon,\delta}}{\epsilon^2}\mathrm{d}t+\frac{g(Z^{\epsilon,\delta})}{\epsilon}\mathrm{d}t=\frac{\sigma\sqrt{2}}{\epsilon\delta}f(Z^{\epsilon,\delta})h(Z^{\epsilon,\delta})\zeta^\delta \mathrm{d}t\\
\mathrm{d}\zeta^\delta=-\frac{\zeta^{\delta}}{\delta^2}\mathrm{d}t+\frac{1}{\delta}\mathrm{d}\beta(t),
\end{cases}
\label{eq:mathsyst}
\end{equation}
where $\bigl(\beta(t)\bigr)_{t\ge 0}$ is a standard real-valued Wiener process (Brownian Motion). For simplicity of the presentation, it is assumed that the initial conditions $Z^{\epsilon,\delta}(0)=z$ and $V^{\epsilon,\delta}(0)=v$ are independent of the parameters $\epsilon$ and $\delta$.

In addition, it is assumed that $\zeta^{\delta}(0)\sim\mathcal{N}(0,1)$ is a centered Gaussian random variable with variance $1$, and is independent of the Wiener process $\beta$. As a consequence, $\bigl(\zeta^\delta(t)\bigr)_{t\ge 0}$ is a stationary Ornstein-Uhlenbeck process: for all $t\ge 0$, $\zeta^\delta(t)\sim\mathcal{N}(0,1)$, and for all $t_1,t_2\ge 0$, the covariance is written as $\mathbb{E}\bigl[\zeta^\delta(t_1)\zeta^\delta(t_2)\bigr]=\frac12 \exp\bigl(-\frac{|t_2-t_1|}{\delta^2}\bigr)$. When $\delta\to 0$, the process $\zeta^\delta$ converges to a white noise, in fact more precisely $\bigl(\frac{1}{\delta}\int_{0}^{t}\zeta^\delta(s)\mathrm{d}s\bigr)_{t\ge 0}$ converges (in distribution) to a Brownian Motion $\bigl(W(t)\bigr)_{t\ge 0}$. However, as will be clear below, one needs to be careful when taking the limit $\delta\to 0$ (in particular concerning the interpretation of the stochastic integral in either It\^o or Stratonovich sense at the limit).

\subsection{Asymptotic expansions}
\label{sec:asymptotic}

The goal of this section is to derive limiting Stochastic Differential Equations for the component $Z^{\epsilon,\delta}$ where the other components are eliminated, when $\epsilon,\delta\to 0$. We will only focus on the derivation of the limiting model, the full rigorous proof of convergence is out of the scope of this work. In this section, the functions $f$, $g$ and $h$ are arbitrary real-valued smooth functions, such that $f(z)>0$ for all $z\in\mathbb{R}$, and with appropriate growth conditions at infinity to ensure global well-posedness of all the SDEs considered below.

\subsubsection{Tools}

A convenient approach \citep{PavliotisStuart} to perform asymptotic analysis in SDEs such as~\eqref{eq:SDE_math} consists in analyzing the behaviour of the associated infinitesimal generator: 
\begin{equation}\label{eq:generator}
\LL^{\epsilon,\delta}=\frac{1}{\epsilon}\A_1+\frac{1}{\epsilon\delta}\A_2+\frac{1}{\epsilon^2}\A_3+\frac{1}{\delta^2}\A_4,
\end{equation}
where, for any smooth function $\varphi:(z,v,\zeta)\in\R^3\mapsto \varphi(z,v,\zeta)\in \R$,
\begin{equation}\label{eq:termes_generator}
\begin{aligned}
\A_1\varphi(z,v,\zeta)&=v\partial_z\varphi(z,v,\zeta)-g(z)\partial_v\varphi(z,v,\zeta),\\
\A_2\varphi(z,v,\zeta)&=\sigma\sqrt{2}h(z)f(z)\zeta \partial_v\varphi(z,v,\zeta),\\
\A_3\varphi(z,v,\zeta)&=-f(z)v\partial_v\varphi(z,v,\zeta)\\
\A_4\varphi(z,v,\zeta)&=-\zeta\partial_\zeta\varphi(z,v,\zeta)+\frac{1}{2}\partial_{\zeta \zeta}^2\varphi(z,v,\zeta).
\end{aligned}
\end{equation}
The second-order differential operator $\LL^{\epsilon,\delta}$ appears on the right-hand side of the backward Kolmogorov equation:
\begin{equation}\label{eq:Kolmogorov}
\begin{cases}
\displaystyle \frac{\partial u^{\epsilon,\delta}(t,z,v,\zeta)}{\partial t}=\LL^{\epsilon,\delta}u^{\epsilon,\delta}(t,z,v,\zeta),\quad t>0,\\
u^{\epsilon,\delta}(0,z,v,\zeta)=u_0(z,v,\eta)
\end{cases}
\end{equation}
for which the solution is given by
\begin{equation}
u^{\epsilon,\delta}(t,z,v,\zeta)=\mathbb{E}_{z,v,\zeta}\Bigl[u_0\Bigl(Z^{\epsilon,\delta}(t),V^{\epsilon,\delta}(t),\zeta^{\epsilon,\delta}(t)\Bigr)\Bigr],
\end{equation}
where the notation $\mathbb{E}_{z,v,\zeta}$ means that the initial conditions are given by $Z^{\epsilon,\delta}(0)=z,V^{\epsilon,\delta}(0)=v,\zeta^{\epsilon,\delta}(0)=\zeta$. By duality, one obtains that the adjoint of the infinitesimal generator $\LL^{\epsilon,\delta}$ is the Fokker-Planck operator, which governs the evolution of the probability density function of the process (Fokker-Planck equation).

The strategy to obtain a limiting SDE for $Z^{\epsilon,\delta}$ consists in the following two steps. First, one identifies the limit of the solution $u^{\epsilon,\delta}$ of the backward Kolmogorov equation~\eqref{eq:Kolmogorov}, for any initial condition $u_0$ which depends only on the $z$ variable. This requires to construct an appropriate asymptotic expansion, to deal with the singular perturbations when $\epsilon , \delta \to 0$. Second, one interprets the limit as the solution of the backward Kolmogorov equation associated with a well-posed SDE. Then one concludes that the limiting model is given by this SDE.

\subsubsection{Limiting Equations}

In the case of small physical parameters, the system of equations Eq.~\ref{eq:mathsyst} converges to a \textit{single} limiting SDE. Three regimes will be studied below:
\begin{enumerate}
\item[Regime 1:] $\epsilon\to 0$, then $\delta\to 0$,
\item[Regime 2:] $\delta\to 0$, then $\epsilon\to 0$,
\item[Regime 3:] $\delta=\lambda\epsilon$, with $\lambda\in(0,\infty)$.
\end{enumerate}
Physically, Regime 1 can be interpreted as $\epsilon \ll \delta \ll 1$, Regime 2 can be interpreted as $\delta \ll \epsilon \ll 1$ and in Regime 3, $\epsilon \sim \delta \ll 1$.

Recall that for Stochastic Differential Equations, the noise may be interpreted either with the It\^o or the Stratonovich convention, and that formulations are equivalent when taking into account a correction term: the It\^o SDE
\begin{equation}
\mathrm{d}X=b(X)\mathrm{d}t+a(X)\mathrm{d}W(t)
\end{equation}
is equivalent to the Stratonovich SDE
\begin{equation}
\mathrm{d}X=\Bigl(b(X)+\frac12a(X)a'(X)\Bigr)\mathrm{d}t+a(X)\circ \mathrm{d}W(t),
\end{equation}
where the notation $a(X)\circ \mathrm{d}W(t)$ is used to precise that the Stratonovich convention is used. The Stratonovich formulation is convenient since it respects the chain rule, whereas for the It\^o formulation one needs to use It\^o's formula. However, the link between an infinitesimal generator, a SDE, and Kolmogorov or Fokker-Planck equations is more clearly seen when using the It\^o formulation. Below, depending on the situation, the most convenient interpretation is chosen.

Below, we prove that the limiting equations are given by the following SDEs:
\begin{enumerate}
\item[Regime 1:] $\mathrm{d}Z=-\frac{g(Z)}{f(Z)}\mathrm{d}t+\sigma\sqrt{2}h(Z)\circ \mathrm{d}W(t)$
\item[Regime 2:] $\mathrm{d}Z=-\frac{g(Z)}{f(Z)}\mathrm{d}t-\frac{\sigma^2 h(Z)^2f'(Z)}{f(Z)}\mathrm{d}t+\sigma\sqrt{2}h(Z)\mathrm{d}W(t)$
\item[Regime 3:] $\mathrm{d}Z=-\frac{g(Z)}{f(Z)}\mathrm{d}t-\frac{\sigma^2 h(Z)(hf)'(Z)}{(1+\lambda^2f(Z))f(Z)}\mathrm{d}t+\sigma\sqrt{2}h(Z)\circ \mathrm{d}W(t)$
\end{enumerate}
where $\bigl(W(t)\bigr)_{t\ge 0}$ is a standard real-valued Wiener process.

Importantly, taking limits $\epsilon\to 0$ then $\delta\to 0$ or $\delta\to 0$ then $\epsilon\to 0$ provides different limiting SDEs. This property originates from stratification. It is not surprising, since if $f$ is a constant function, then the It\^o formulation of the SDE of Regime 2 gives $\mathrm{d}Z=-\frac{g(Z)}{f(Z)}\mathrm{d}t+\sigma\sqrt{2}\frac{h(Z)}{f(Z)}\mathrm{d}W(t)$: the SDEs of Regime 1 and Regime 2 differ by an It\^o-Stratonovich correction term. However, this observation does not hold if $f$ is not constant: indeed the It\^o formulation of the SDE of Regime 2 is
\begin{equation}
\mathrm{d}Z=-\frac{g(Z)}{f(Z)}\mathrm{d}t-\frac{\sigma^2 h^2(Z)f'(Z)}{f^3(Z)}\mathrm{d}t+\sigma\sqrt{2}h(Z)\mathrm{d}W(t).
\end{equation}
More precisely, consider the case $h=1$, with a non-constant $f$ (this is the most important case in this study). Whereas It\^o and Stratonovich interpretations coincide, the limiting SDEs differ by the presence of an additional {\it noise-induced drift} term (\citealt{Hottovy12,Hottovy15,Herzog16,FreidlinHu}) Observe that, formally, Regime 1 (resp. Regime 2) corresponds to Regime 3 when $\lambda=\infty$ (resp. $\lambda=0$).

The physical consequence of this result is that one has to be extremely careful when choosing the Regime to interpret the dynamics of the system (see Appendix~\ref{sec:deriv} for detailed calculations).

\subsection{Steady-state dust distributions}
\label{sec:analytic_distribs}

\subsubsection{Constant diffusivity}

In this section, it is assumed that $h=1$ is a constant. As a consequence, It\^o and Stratonovich interpretations of the limiting SDEs coincide, since the diffusion coefficient is constant. However, stratification means that $f$ is not constant, thus a noise-induced drift term appears. With the convention that Regime 1 (resp. Regime 2) is obtained with $\lambda=\infty$ (resp. $\lambda=0$), the limiting SDE is written as
\begin{equation}
\mathrm{d}Z=-\frac{g(Z)}{f(Z)}\mathrm{d}t-\frac{\sigma^2 f'(Z)}{(1+\lambda^2f(Z))f(Z)}\mathrm{d}t+\sigma\sqrt{2}\mathrm{d}W(t).
\label{eq:SDE_lim}
\end{equation}

This SDE is rewritten as the overdamped Langevin equation
\begin{equation}
\mathrm{d}Z=-\nabla\mathcal{V}_\lambda^{\sigma}(Z)\mathrm{d}t+\sigma \sqrt{2}\mathrm{d}W(t),
\end{equation}
where the potential energy function $\mathcal{V}_\lambda^\sigma$ is defined as
\[
\mathcal{V}_\lambda(z)=\mathcal{V}_\infty(z)+ \mathcal{V}_{\rm corr}(z) .
\]
$\mathcal{V}_\infty$ denotes the antiderivative of $g/f$ and $\mathcal{V}_{\rm corr}$ the antiderivative of $\sigma^{2} f'/\left(\left[ 1 + \lambda^{2} f \right] f \right)$, i.e. 
\[
\mathcal{V}_{\rm corr} =  \sigma^2\log\bigl(\frac{f(z)}{1+\lambda^2f(z)}\bigr).
\]

As a consequence, under appropriate conditions on the growth at infinity of $\mathcal{V}_{\lambda}^{\sigma}$ (which are satisfied in the example considered below), the limiting SDE defines an ergodic dynamics, with unique invariant distribution having the density
\begin{equation}
\rho_\lambda^\sigma(z)=\frac{1}{Z_\lambda^\sigma}\exp\bigl(-\frac{\mathcal{V}_\lambda^{\sigma}(z)}{\sigma^2}\bigr),
\label{eq:pdf_steady}
\end{equation}
with normalization constant $Z_\lambda^\sigma=\int_{-\infty}^{+\infty}e^{-\frac{\mathcal{V}_\lambda^\sigma(z)}{\sigma^2}}\mathrm{d}z$.

\bigskip

The parameters $\lambda$ and $\sigma$ may considerably change the qualitative properties of the potential energy function $\mathcal{V}_\lambda^\sigma$. For instance, choose the functions $f$ and $g$ as follows:
\[
f(z)=e^{-\frac{z^2}{2}}~,\quad g(z)=z,
\]
which gives $\mathcal{V}_\infty(z)=e^{\frac{z^2}{2}}$. Observe that this potential energy function is convex, with a unique global minimum located at $z=0$. However, straightforward computations give
\[
\nabla \mathcal{V}_\lambda^\sigma(0)=0~,\quad \nabla^2 \mathcal{V}_\lambda^\sigma(0)=\frac{1}{2}-\frac{\sigma^2}{ 1 + \lambda^{2}},
\]
thus $0$ is not a minimum of $\mathcal{V}_\lambda^\sigma$ if $\sigma^2>1+\lambda^2$. Hence, the steady dust density can either be single- or double-hump shaped. For eddy times of order unity and typical disc parameters, the asymptotic distribution obtained from Eq.~\ref{eq:pdf_steady} does not differ much from the model of \citet{Fromang2009}. This is not the case anymore in the diffusive limit $t_{\rm e} \to 0$.

\subsubsection{Stratified diffusivity}

We address vertical gradients of the  diffusivity $\alpha$ via the simple parametrisation $h\left(z \right) = f(z)^{-1/4}$ discussed in Sect.~\ref{sec:lagturb}. We obtain
\begin{align}
\mathcal{V}_{\rm corr}(z) & = \sigma^{2} \int \frac{h \left(fh \right)'}{\left(1 + \lambda^{2} f \right) f} \\
& = - \frac{3 \sigma^{2}}{4}\left\lbrace \frac{2}{\sqrt{f}} + 2 \lambda \tan^{-1}\left( \lambda \sqrt{f} \right) . \right\rbrace .
\end{align}
Hence, 
\[
\nabla \mathcal{V}_\lambda^\sigma(0)=0~,\quad \nabla^2 \mathcal{V}_\lambda^\sigma(0)=\frac{1}{2}-\frac{3\sigma^2}{8\left( 1 + \lambda^{2} \right)}.
\]
$0$ is therefore not a minimum of $\mathcal{V}_\lambda^\sigma$ if $\sigma^2>\frac{4\left( 1+\lambda^2\right)}{3}$. Except for a marginal set of nonphysical parameters, dust distributions that account form the vertical dependency of the diffusivity are almost similar to the one obtained for constant values of $\alpha$. Stochastic turbulent driving scales as $\sqrt{\alpha} / \rho_{\rm g} \propto hf = f^{3/4}$ for our model, hence preserving the essential of the $h = 1$ settling mechanism.

\subsubsection{On the development of bumps in the diffusive limit}

\begin{figure}
  \includegraphics[width=.45 \textwidth]{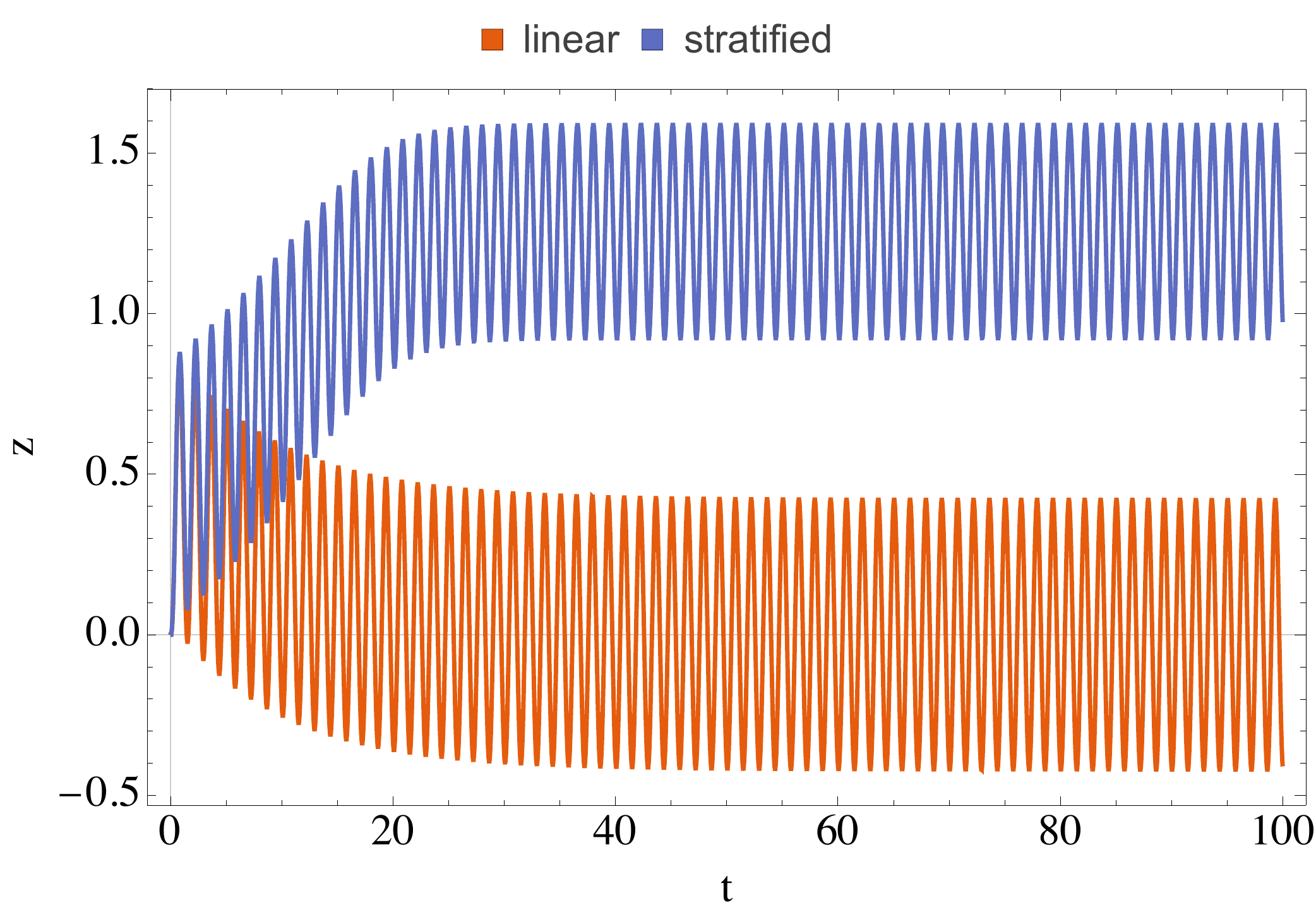}
   \caption{Evolution of dust grains driven by toy sinusoidal gas velocities. Lift-up of small grains resulting from the differential driving that originates form stratification (solid blue line). For an homogeneous disc, grains relax and oscillate around the midplane (solid red line). We adopt $S_{\! \rm t 0} = 0.1$, $G = 2$, $\omega = 4.4$ and $z_{0} = \dot{z}_{0} = 0$.}
   \label{fig:oscill}
\end{figure}

A striking feature of the asymptotic distributions obtained in Sect.~\ref{sec:analytic_distribs} is the development of dust over-concentrations above the midplane in the limit $t_{\rm e} \to 0$. Fig.~\ref{fig:oscill} corroborates this finding by comparing the evolution of the two following oscillators
\begin{equation}
\ddot{z} + S_{\! \rm t 0}^{-1}  \mathrm{e}^{-z^2/2} \dot{z} + z = G \, S_{\! \rm t 0}^{-1} \mathrm{e}^{-z^2/2} \sin \left( \omega t \right) ,
\label{eq:oscill_nl}
\end{equation}
and its linearised version
\begin{equation}
\ddot{z} + S_{\! \rm t 0}^{-1}   \dot{z} + z  = G \, S_{\! \rm t 0}^{-1} \sin \left( \omega t \right) ,
\label{eq:oscill_lin}
\end{equation}
for $S_{\! \rm t 0} = 0.1$, $G = 2$, $\omega = 4.4$ and $z_{0} = \dot{z}_{0} = 0$ (those parameters are chosen to make the figure clear). $G$ and $\omega$ parametrise the intensity and the frequency of the driving and play the role of $\alpha$ and $\tau_{\rm e}$ in the stochastic model. Fig.~\ref{fig:oscill} shows spontaneous symmetry breaking between the top and the bottom layers of the stratified disc. Physically, the lift-up of small grains results from i) an important inertia when grains reach the top layers of the disc, ii) a modulated intensity of the turbulent driving by stratification that sets the maximum gradient of turbulent driving at one pressure scale height, and iii) a driving frequency that is large enough for this differential effect to cumulate. This is always the case when $\omega \to \infty$, which corresponds to $\tau_{\rm e}\to 0$. Hence, grains are constantly kicked from below by the differential driving and are lifted up above the midplane, explaining the formation of the dusty bumps.

\section{Numerical results}
\label{sec:numerics}

\subsection{Numerical scheme}
\label{sec:scheme}

We now aim to validate Eq.~\ref{eq:pdf_steady}, i.e. the formula obtained for the invariant distribution of the limiting SDE by direct numerical simulation of Eqs.~\ref{eq:sto1},\ref{eq:sto2},\ref{eq:sto3}. When changing the parameters, we illustrate the apparition of double-humped shaped instead of single-humped distributions. Eqs.~\ref{eq:sto1},\ref{eq:sto2},\ref{eq:sto3} are solved numerically with a Strang splitting method, observing that the sub-systems
\begin{eqnarray}
\mathrm{d}Z & = & 0  , \label{eq:split11}\\
\mathrm{d}V + S_{\! \rm t 0}^{-1}  f_{\phi}(Z) V \mathrm{d}T + g_{\phi}(Z) \mathrm{d}T  & = & S_{\! \rm t 0}^{-1} f_{\phi}(Z) h(Z) \sqrt{2\alpha_{0}}  \,  \xi\mathrm{d}T, \label{eq:split12} \\
\mathrm{d}\xi & = & 0, \label{eq:split13}
\end{eqnarray}
and
\begin{eqnarray}
\mathrm{d}Z & = & V \mathrm{d}T  , \label{eq:split21}\\
\mathrm{d}V& = & 0 ,\label{eq:split22}\\
\mathrm{d}\xi & = & -\frac{\xi}{\tau_{\rm e}}\mathrm{d}T + \frac{\mathrm{d} w}{\tau_{\rm e}}, \label{eq:split23}
\end{eqnarray}
can be solved exactly. On the one hand, the solution at any time $T>0$ of the system of Eqs.~\ref{eq:split11},\ref{eq:split12},\ref{eq:split13} is given by
\begin{equation}
\Psi_t^{(1)}(Z_0,V_0,\xi_0)=
\begin{cases}
Z_0,\\
e^{-\frac{f_\phi(Z_0)T}{S_{\! \rm t 0}}}V_0+\\
\left(1-\mathrm{e}^{-\frac{f_\phi(Z_0)T}{S_{\! \rm t 0}}}\right)\Bigl(h(Z_0) \sqrt{2\alpha_{0}}  \,  \xi_0-\frac{S_{\! \rm t 0}}{f_\phi(Z_0)}g_\phi(Z_0)\Bigr),\\
\xi_0.
\end{cases}
\end{equation}
On the other hand, the solution at any time $T>0$ of the system of Eqs.~\ref{eq:split21},\ref{eq:split22},\ref{eq:split23} is given by
\begin{equation}
\Psi_t^{(2)}(Z_0,V_0,\xi_0)=\begin{cases}
Z_0+TV_0,\\
V_0,\\
\mathrm{e}^{-\frac{T}{\tau_{\rm e}}}\xi_0+\frac{1}{\tau_{\rm e}}\int_{0}^{T}\mathrm{e}^{-\frac{T-T'}{\tau_{\rm e}}}\mathrm{d}w(T'),
\end{cases}
\end{equation}
where $\frac{1}{\tau_{\rm e}}\int_{0}^{t}\mathrm{e}^{-\frac{t-s}{\tau_{\rm e}}}\mathrm{d}w(s)\sim\mathcal{N}\left(0,\frac{1}{2\tau_{\rm e}}(1-\mathrm{e}^{-\frac{2T}{\tau_{\rm e}}})\right)$ follows a Gaussian distribution. Given a time-step size $\Delta T>0$, then the Strang splitting scheme is defined by the recursion
\begin{equation}
\left(Z_{n+1},V_{n+1},\xi_{n+1}\right)=\left(\Psi_{\frac{\Delta T}{2}}^{(2)}\circ \Psi_{\Delta t}^{(1)}\circ \Psi_{\frac{\Delta T}{2}}^{(2)}\right)\left(Z_n,V_n,\xi_n\right),
\end{equation}
\begin{figure*}
	\centering
	\includegraphics[width=\columnwidth]{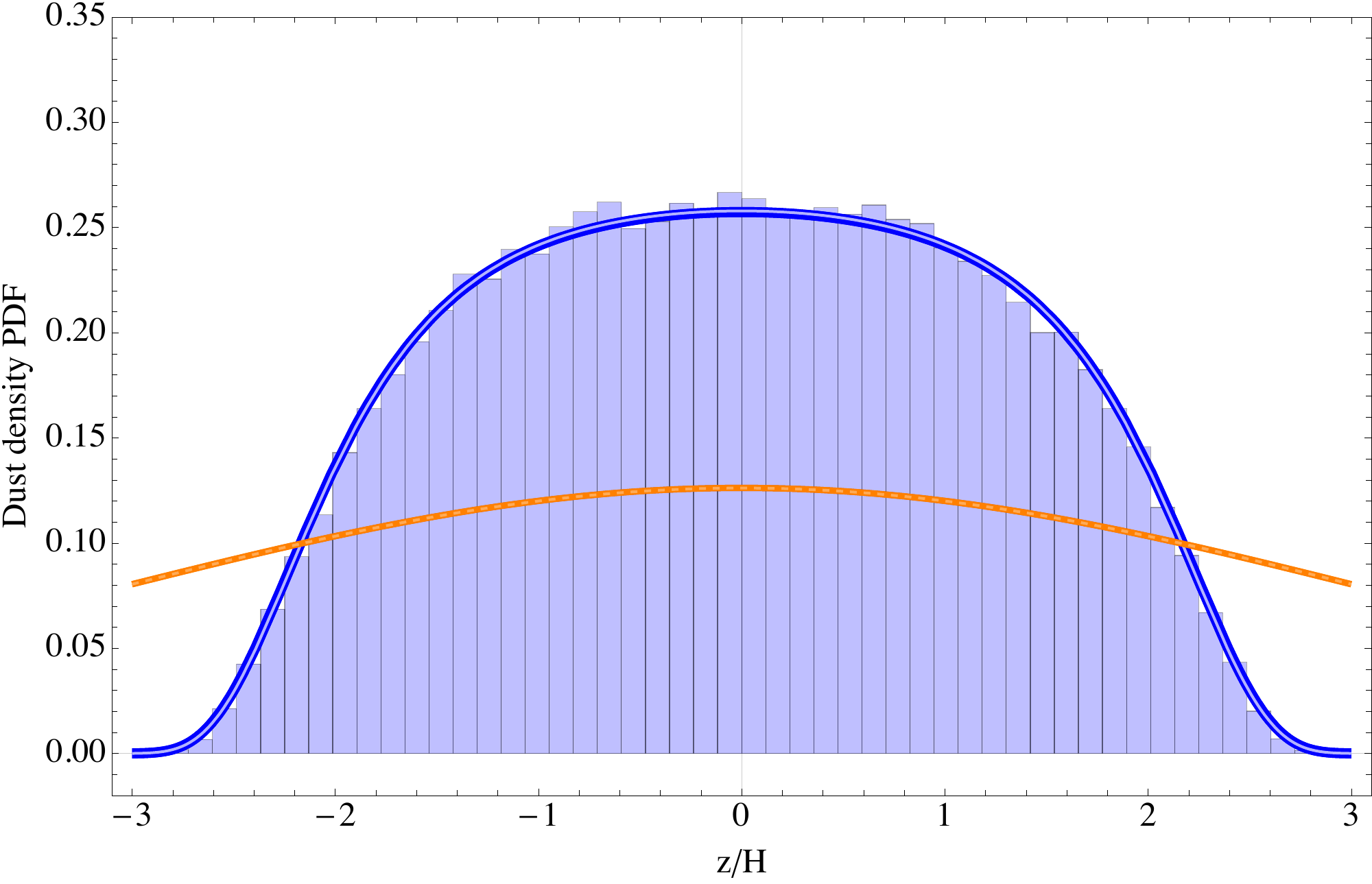}
    \includegraphics[width=\columnwidth]{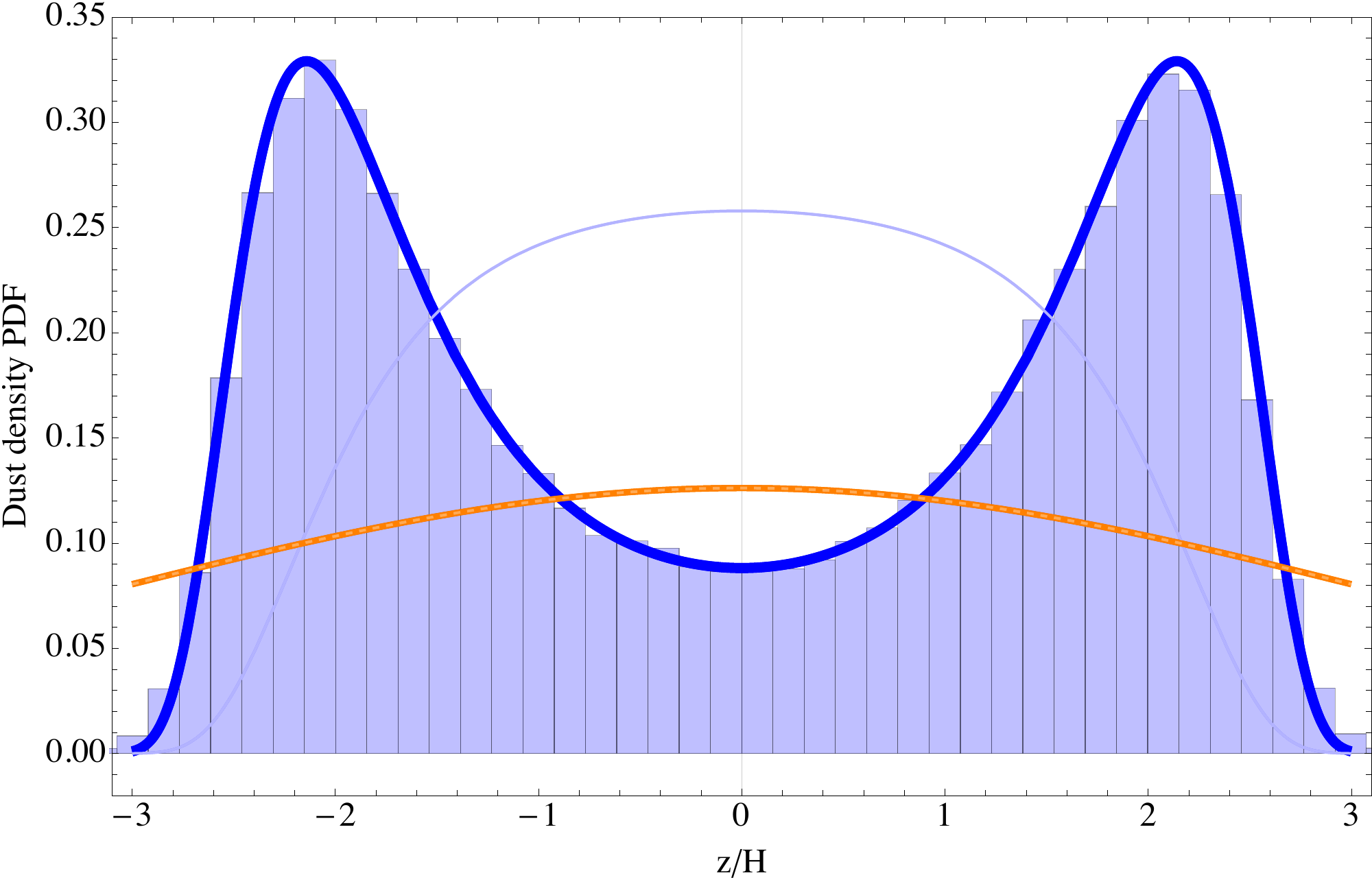}
	\caption{Left: Histogram obtained for $10^{5}$ particles in a configuration where $\alpha = 0.01$, $S_{\mathrm{t},0} = 0.001$ and $\tau_{\rm e} = 1$ after a time $t = 10^{3}$. The asymptotic solution obtained in Sect.~\ref{sec:asymptotic} is represented by the solid thick blue line. Other curves represent predictions from the \citet{Dubrulle1995} model (dashed light blue line), the \citet{Youdin2007} model (dot-dashed lighter blue line) and the \citet{Fromang2009} model (solid lightest blue line). The steady-state distribution is single-humped and is flatter than a Gaussian, as predicted by our model and the one of \citet{Fromang2009}. Right: Similar plot, but for $\alpha = 0.1$, $S_{\mathrm{t},0} = 0.01$ and $\tau_{\rm e} = 0.001$. This case corresponds to the purely diffusive limit, and grains are found to over-concentrate well above the midplane of the disc. This double-hump shape is recovered only by our asymptotic expansion. }
	\label{fig:num}
\end{figure*}
and each step is made of three succesive updates. Let $\gamma_{0,1},\gamma_{0,2},\ldots,\gamma_{n,1},\gamma_{n,2},\ldots$ be independent $\mathcal{N}(0,1)$ standard Gaussian random variables. First, using the definition of $\Psi_{\frac{\Delta T}{2}}^{(2)}$, and a random variable $\gamma_{n,1}\sim\mathcal{N}(0,1)$, let
\begin{equation}
\begin{cases}
Z_n\leftarrow Z_n+\frac{\Delta T}{2}V_n,\\
V_n\leftarrow V_n,\\
\xi_n\leftarrow \mathrm{e}^{-\frac{\Delta T}{2\tau_{\rm e}}}\xi_n+\frac{1}{\sqrt{2\tau_{\rm e}}}\sqrt{1-\mathrm{e}^{-\frac{\Delta T}{\tau_{\rm e}}}}\gamma_{n,1}.
\end{cases}
\end{equation}
Second, using the definition of $\Psi_{\Delta T}^{(1)}$, let
\begin{equation}
\begin{cases}
Z_n\leftarrow Z_n,\\
V_n\leftarrow \mathrm{e}^{-\frac{f_\phi(Z_n)\Delta T}{S_{\! \rm t 0}}}V_n+\left(1-\mathrm{e}^{-\frac{f_\phi(Z_n)\Delta T}{S_{\! \rm t 0}}}\right)\Bigl(h(Z_n) \sqrt{2\alpha_{0}}  \,  \xi_n-\frac{S_{\! \rm t 0}}{f_\phi(Z_n)}g_\phi(Z_n)\Bigr),\\
\zeta_n\leftarrow \zeta_n.
\end{cases}
\end{equation}
Using the definition of $\Psi_{\frac{\Delta T}{2}}^{(1)}$ and a random variable $\gamma_{n,2}\sim\mathcal{N}(0,1)$,
\begin{equation}
\begin{cases}
Z_n\leftarrow Z_n+\frac{\Delta T}{2}V_n,\\
V_n\leftarrow V_n,\\
\xi_n\leftarrow \mathrm{e}^{-\frac{\Delta T}{2\tau_{\rm e}}}\xi_n+\frac{1}{\sqrt{2\tau_{\rm e}}}\sqrt{1-\mathrm{e}^{-\frac{\Delta T}{\tau_{\rm e}}}}\gamma_{n,2},
\end{cases}
\end{equation}
and one sets
\begin{equation}
Z_{n+1}\leftarrow Z_n~,\quad V_{n+1}\leftarrow V_n~,\quad \xi_{n+1}\leftarrow \xi_n.
\end{equation}

\subsection{Numerical dust distributions}
\label{sec:distribs}

 We adopt a Courant-Friedrich-Levy condition of $\Delta t \propto S_{\! \rm t 0} \ll 1$ and use a safety factor of $0.1$ gathered from a numerical convergence analysis. The probability density distributions reach steady-state for $ t \sim S_{\! \rm t 0}^{-1}$, the settling time of small dust grains. Fig.~\ref{fig:num} shows histograms obtained for $10^{5}$ particles initially placed in the midplane with no velocity. In this configuration, sufficient accuracy is obtained to validate the model. Our first simulation consists of a seminal disc with $\alpha = 0.01$ and $\tau_{\rm e} = 1$, populated with small grains with Stokes number in the midplane $S_{\mathrm{t},0} = 10^{-3}$. Fig.~\ref{fig:num} (left) shows that the steady-state distribution is correctly reproduced by our asymptotic description and the \citet{Fromang2009} model, the two curves being nearly superimposed in this regime. In particular, flatter distributions than Gaussian are obtained. Stratification gradients push more grains from the midplane to the top layers of the disc than in an homogeneous configuration. Almost no grains above $z = 3$ are found. This is expected as there is almost no gas at this height and dust grains settle back to the midplane until they got stuck again. Our second simulation is designed to demonstrate the accuracy of our asymptotic expansion. We setup an academic configuration where $\alpha = 0.1$, $S_{\mathrm{t},0} = 10^{-2}$ and $\tau_{\rm e} = 10^{-3}$ to reach the purely diffusive limit while preserving numerical tractability. Fig.~\ref{fig:num} (right) shows that again, the steady-state distribution is correctly reproduced by our asymptotic expansion. In an obvious manner, the double-hump shape with strong over-concentrations of dust at $z \simeq 2$ is correctly captured. Alternative models predict incorrect bell-shaped distributions in this regime. In this regime, the rate of differential kicks received by the grains is extremely important and the cumulative contribution powers up the lift-up of the particles.
 
 Importantly, these peaks can arise as a parasitic effect when equations of motion are integrated with stratification, inertia, but in the diffusive limit with zero eddy-time. Hence the necessity of integrating the settling equations including a finite turbulent timescales.
\begin{figure}
	\includegraphics[width=\columnwidth]{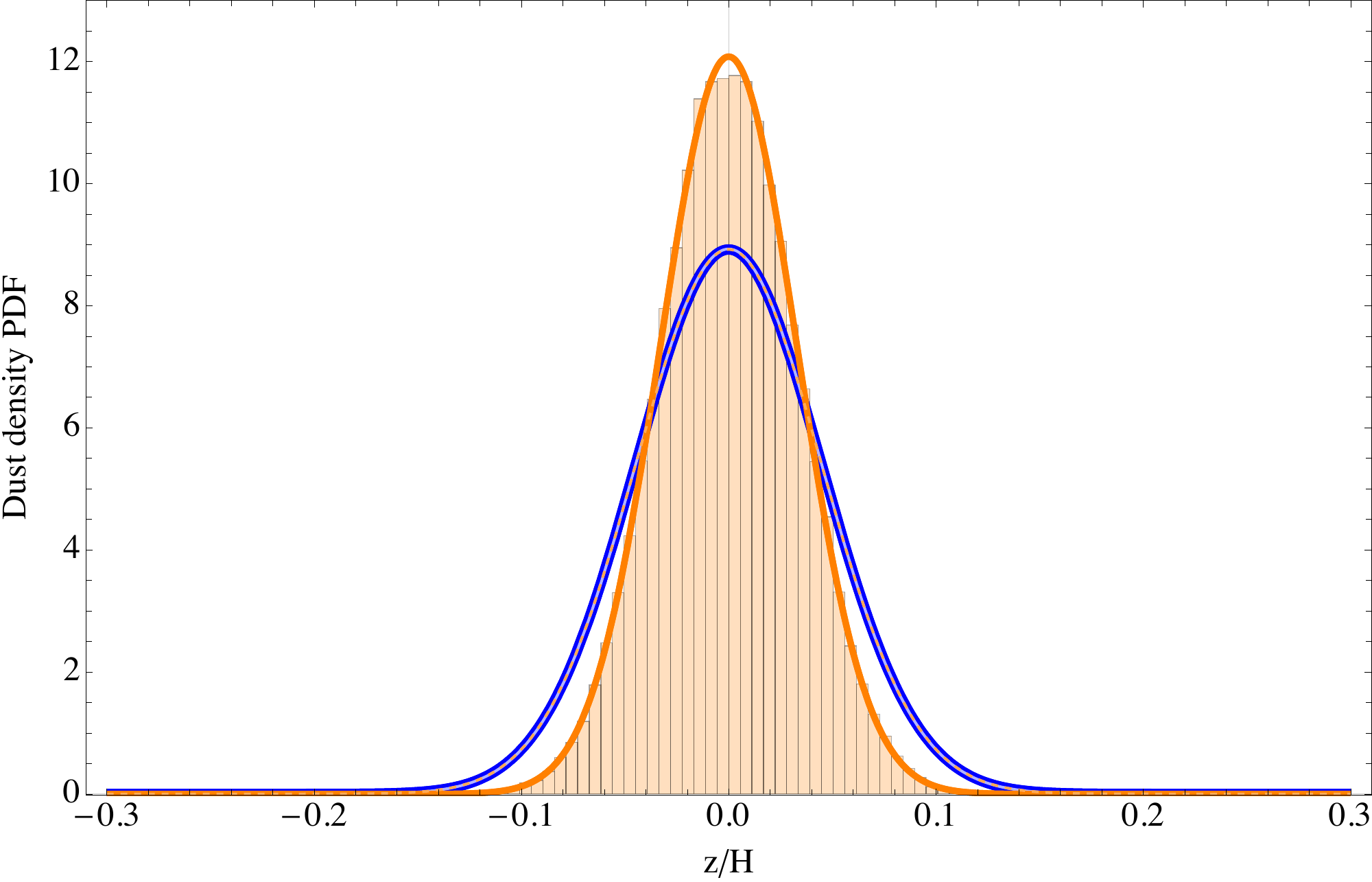}
	\caption{Similar to Fig.~\ref{fig:num}, but with $\alpha = 0.01$, $S_{\mathrm{t},0} = 5$ and $\tau_{\rm e} = 1$, a regime where our asymptotic expansion is not valid. The model of \citet{Youdin2007} is the most accurate in this regime.}
	\label{fig:youdin}
\end{figure}
Finally, it should be noted that our asymptotic expansion does not hold for large grains $S_{\mathrm{t},0} \gtrsim 1$ that remain close to the midplane. Fig.~\ref{fig:youdin} shows that in that case, best accuracy is obtained by the \citet{Youdin2007} model. Interestingly, finite eddy time terms gives noticeable corrections in this case as well.

\section{Discussion}
\label{sec:discussion}

The model of disc considered in this study remains fiducial. Gas does not undergo any dynamical evolution such as outflows, winds, viscous spreading or evaporation. We did not consider gravitating bodies embedded in the disc and have restrained the study to grains of constant size that neither grow nor fragment. We also focused on steady-state distributions, since they are widely used in as practical recipes for dust densities. As a short remark on this point, we note that steady-state is reached after a few settling times (Eqs.~\ref{eq:sto1},\ref{eq:sto2}). For small grains, this time is orders of magnitude longer than other dynamical times in the disc. We put therefore a strong warning against using these formulae in vertically integrated models, to estimate instantaneous volume concentrations from surface densities. Finally, we note that dust lift-up may become significant in stratified objects that are trans- or supersonic and contain small grains, such as molecular clouds. In this case, dust may be lifted up easily by turbulence even when it develops on large integral timescales, as long as the cloud remains stable over a time that is sufficiently long for the grains to differentiate spatially.

\section{Conclusion}
\label{sec:conclu}

In the context of better understanding observations of small dust grains in young discs, we derived refined analytic prescriptions for the distributions of small grains that populate their top layers. Our model includes  gas stratification, dust inertia and finite correlation times for the turbulence. It is derived from first principles, by writing a balance of forces on a grain where stochastic driving mimics rigorously the statistical properties of homogeneous isotropic turbulence. The role of the vertical gradient of $\alpha$ is investigated using the scaling $\alpha \propto \rho^{-1/2}$, which ensures a constant density of turbulent energy through the disc. From rigorous asymptotic expansions, we obtained steady-state distributions for small grains scattered through the stratified disc by turbulence. Unexpected technicalities arise to handle small Stokes numbers in the diffusive limit. These novel distributions are validated against a direct numerical integration of the stochastic system via a Strang-splitting scheme. The main results of this study are summarised below.
\begin{enumerate}
\item Let consider a disc orbiting with frequency $\Omega$ and gas scale height $H$, for which the turbulent activity and the lifetime of the largest eddies are parametrised by $\alpha$ and $t_{\rm e}$. We consider grains that have Stokes numbers in the midplane smaller than unity, i.e. $S_{\! \rm t 0}~\ll~1$. Dust density a steady-state is given by
\begin{equation}
\rho_{\rm d}(z) \propto \left( \frac{t_{\rm e}\Omega}{S_{\! \rm t 0}} +\mathrm{e}^{\frac{z^{2}}{2 H^{2}}} \right)\exp\left(-\frac{\mathrm{e}^{ \frac{z^{2}}{2 H^{2}}}}{\alpha / S_{\! \rm t 0} }\right),
\end{equation}
which corresponds to Eq.~\ref{eq:pdf_steady} expressed in physical quantities.

\item For $t_{\rm e} \sim \Omega^{-1}$, dust distributions are single-humped and flattened. In the purely diffusive limit $t_{\rm e} \ll \Omega^{-1}$, dust distributions become bumpy and develop non-physical strong peaks at $z \ge H$. As such, turbulent correlations must be handled with care in settling models.

\item Observations support the absence of dust over-concentrations above the scale height of young disc, hence corroborating numerical experiments predicting $t_{\rm e} \sim \Omega^{-1}$.

\end{enumerate}
Trans- or supersonic stratified systems such as molecular clouds may enter the regime of parameters where dust lift-up may becomes important and should deserve further investigations.

\section*{Acknowledgments}
G. Laibe thanks S. Fromang and G. Lesur, who first pointed out the role played by finite correlation times in this problem. We also thank P. Sandquist, P. Woitke, I. Bonnell, and the referee for his/her report. This project was supported by the IDEXLyon project (contract nANR-16-IDEX-0005) under the auspices University of Lyon. We acknowledge financial support from the national programs (PNP, PNPS, PCMI) of CNRS/INSU, CEA, and CNES, France. This project has received funding from the European Union's Horizon 2020 research and innovation program under the Marie Sk\l odowska-Curie grant agreement No 823823.

\begin{appendix}
\section{Derivation of the limiting equations}

\subsection{Analysis in Regime 1}

The derivation of the limiting SDE in Regime 1 follows from standard arguments and does not contain any difficulty or unexpected additional term. We thus only provide the heuristic arguments. A rigorous analysis may be performed using the tools developed below to deal with the other regimes.

For the first step, the parameter $\delta>0$ is held fixed, and  one needs to pass to the limit $\epsilon\to 0$. Observe that
\begin{equation}
\begin{aligned}
\mathrm{d}Z^{\epsilon,\delta}&=\frac{V^{\epsilon,\delta}}{\epsilon}\mathrm{d}t\\
&=-\frac{g(Z^{\epsilon,\delta})}{f(Z^{\epsilon,\delta})}\mathrm{d}t+\frac{\sigma\sqrt{2}}{\delta}h(Z^{\epsilon,\delta})\zeta^\delta \mathrm{d}t-\epsilon \mathrm{d}V^{\epsilon,\delta},
\end{aligned}
\end{equation}
and as a consequence the limiting SDE when $\epsilon\to 0$ is given by
\begin{equation}
\begin{cases}
\mathrm{d}Z^{0,\delta}=-\frac{g(Z^{0,\delta})}{f(Z^{0,\delta})}\mathrm{d}t+\frac{\sigma\sqrt{2}}{\delta}h(Z^{0,\delta})\zeta^\delta \mathrm{d}t,\\
\mathrm{d}\zeta^\delta=-\frac{\zeta^{\delta}}{\delta^2}\mathrm{d}t+\frac{1}{\delta}\mathrm{d}\beta(t).
\end{cases}
\end{equation}

With the notation $\eta^\delta(t)=\delta^{-1}\int_0^t\zeta^\delta(s)\mathrm{d}s$, one has
\begin{equation}
\frac{1}{\delta}\zeta^\delta \mathrm{d}t=\mathrm{d}\eta^\delta=\mathrm{d}\beta(t)-\delta \mathrm{d}\zeta^\delta,
\end{equation}
which heuristically justifies convergence of $\eta^\delta$ to Brownian Motion. At the limit, noise needs to be interpreted with the Stratonovich convention, which is a classical result when Brownian Motion is approximated by a smooth process. Thus, passing to the limit $\delta\to 0$, one obtains the limit SDE
\begin{equation}
\mathrm{d}Z=-\frac{g(Z)}{f(Z)}\mathrm{d}t+\sigma\sqrt{2}h(Z)\circ \mathrm{d}W(t),
\end{equation}
where $\bigl(W(t)\bigr)_{t\ge 0}$ is a real-valued standard Wiener process.

The equivalent It\^o formulation of the SDE is
\begin{equation}
\mathrm{d}Z=-\frac{g(Z)}{f(Z)}\mathrm{d}t+\sigma^2h(Z)h'(Z)\mathrm{d}t+\sigma\sqrt{2}h(Z)\mathrm{d}W(t).
\end{equation}

\subsection{Analysis in Regime 2}

In this regime, one needs to be careful in order to exhibit the noise-induced drift term when $f$ is not constant. We thus provide all the details of the derivation.

Note that the first step below still follows from a standard argument (which is made rigorous below): for fixed $\epsilon>0$, when $\delta\to 0$, one obtains the limiting SDE
\begin{equation}\label{eq:SDE_intermediaire}
\begin{cases}
\mathrm{d}Z^\epsilon=\frac{V^\epsilon}{\epsilon}\mathrm{d}t\\
\mathrm{d}V^\epsilon+\frac{f(Z^{\epsilon})V^{\epsilon}}{\epsilon^2}\mathrm{d}t+\frac{g(Z^{\epsilon})}{\epsilon}\mathrm{d}t=\frac{\sigma\sqrt{2}}{\epsilon}f(Z^{\epsilon})h(Z^{\epsilon})\mathrm{d}\tilde{W}(t),
\end{cases}
\end{equation}
where $\bigl(\tilde{W}(t)\bigr)_{t\ge 0}$ is a real-valued standard Wiener process. Observe that It\^o and Stratonovich interpretations of the noise coincide for this SDE (the diffusion coefficient depends only on the position component, whereas the noise acts only on the velocity component). However, an heuristic argument to pass to the limit $\epsilon\to 0$ would not explain the presence of the noise-induced drift term (when $f$ is not constant), and thus would not provide the correct limiting SDE.

Let us now present a rigorous derivation of the limiting SDE in Regime 2. For the first step, the parameter $\epsilon>0$ is held fixed. One needs to construct an asymptotic expansion in terms of the small parameter $\delta$, of the form
\begin{equation}
u^{\epsilon,\delta}(t,z,v,\zeta)=u^{\epsilon,0}(t,z,v)+\delta r^{\epsilon,1}(t,z,v,\zeta)+\delta^2 r^{\epsilon,2}(t,z,v,\zeta)+{\rm O}(\delta^3),
\end{equation}
where the zero-order term $u^{\epsilon,0}$ does not depend on $\zeta$ and describes the limiting process. Then one needs to identify the limiting generator $\overline{\LL}^{\epsilon,0}$ such that one has $\partial_tu^{\epsilon,0}=\overline{\LL}^{\epsilon,0}u^{\epsilon,0}$.

Inserting the asymptotic expansion in the backward Kolmogorov equation~\eqref{eq:Kolmogorov} and using the expression~\eqref{eq:generator} of the infinitesimal generator $\LL^{\epsilon,\delta}$, one obtains the following hierarchy of equations when matching terms of size $\delta^{-2}$, $\delta^{-1}$ and $1$ respectively:
\begin{equation}
\begin{aligned}
\A_4u^{\epsilon,0}&=0,\\
\A_4r^{\epsilon,1}+\frac{1}{\epsilon}\A_2u^{\epsilon,0}&=0,\\
\A_4r^{\epsilon,2}+\frac{1}{\epsilon}\A_2r^{\epsilon,1}+\bigl(\frac{1}{\epsilon}\A_1+\frac{1}{\epsilon^2}\A_3\bigr)u^{\epsilon,0}&=\partial_tu^{\epsilon,0}.
\end{aligned}
\end{equation}
The first equation is consistent with the assumption that $u^{\epsilon,0}$ does not depend on $\zeta$. A solution of the second equation is given by
\begin{equation}
r^{\epsilon,1}(t,z,v,\zeta)=\frac{\sigma\sqrt{2}f(z)h(z)}{\epsilon}\zeta \partial_v u^{\epsilon,0}(t,z,v).
\end{equation}

Let $\nu=\mathcal{N}(0,\frac12)$ denote the invariant distribution of the Ornstein-Uhlenbeck process $\mathrm{d}\zeta=-\zeta \mathrm{d}t+\mathrm{d}\beta(t)$. The partial differential equation satisfied by $u^{\epsilon,0}$ is obtained by taking the average of the last equation of the hierarchy, with respect to $\mathrm{d} \nu(\zeta)$, and using the property $\int \mathcal{A}_4\psi(\zeta)\mathrm{d}\nu(\zeta)=0$, for any smooth function $\psi$. Using that $\int \zeta^2 \mathrm{d}\nu(\zeta)=\frac12$, one obtains
\begin{equation}
\begin{aligned}
\partial_tu^{\epsilon,0}&=\frac{\sigma^2f(z)^2h(z)^2}{\epsilon^2}\partial_{vv}^2 u^{\epsilon,0}+\bigl(\frac{1}{\epsilon}\A_1+\frac{1}{\epsilon^2}\A_3\bigr)u^{\epsilon,0}\\
&=\overline{\LL}^{\epsilon,0}u^{\epsilon,0}.
\end{aligned}
\end{equation}
For completeness, $r^{\epsilon,2}$ is constructed as solution of the Poisson equation
\begin{equation}
-\A_4r^{\epsilon,2}(t,z,v,\zeta)=\frac{1}{\epsilon}\bigl(\A_2r^{\epsilon,1}(t,z,v,\zeta)-\int \A_2r^{\epsilon,1}(t,z,v,\cdot)\mathrm{d}\nu\bigr),
\end{equation}
which is solvable since the right-hand side is centered with respect to $\nu$.

The limiting generator $\overline{\LL}^{\epsilon,0}$ is associated with the SDE~\eqref{eq:SDE_intermediaire}. It remains now to pass to the limit $\epsilon\to 0$. This is performed by constructing an asymptotic expansion in terms of the small parameter $\epsilon$ of the form
\begin{equation}
u^{\epsilon,0}(t,z,v)=u^{0,0}(t,z)+\epsilon r^{0,1}(t,z,v)+\epsilon^2 r^{0,2}(t,z,v)+{\rm O}(\epsilon^3),
\end{equation}
where the zero-order term $u^{0,0}$ does not depend on $v$, and by identifying the limiting generator $\overline{\LL}^{0,0}$ such that one has $\partial_t u^{0,0}=\overline{\LL}^{0,0}u^{0,0}$. Observe that one can write
\begin{equation}
\overline{\LL}^{\epsilon,0}=\frac{1}{\epsilon}\A_1+\frac{1}{\epsilon^2}\overline{\A}_3,
\end{equation}
where $\overline{\A}_3$ is defined by
\begin{equation}
\overline{\A}_3\varphi(z,v)=\sigma^2 f(z)^2 h(z)^2\partial_{vv}^2\varphi(z,v)-f(z)v\partial_v\varphi(z,v).
\end{equation}

Inserting the asymptotic expansion in the backward Kolmogorov equation yields the following hierarchy of equations, when matching terms of size $\epsilon^{-2}$, $\epsilon^{-1}$ and $1$ respectively:
\begin{equation}
\begin{aligned}
\overline{\A}_3u^{0,0}&=0,\\
\overline{\A}_3r^{0,1}+\A_1u^{0,0}&=0,\\
\overline{\A}_3r^{0,2}+\A_1r^{0,1}&=\partial_tu^{0,0}.
\end{aligned}
\end{equation}
The first equation is consistent with the assumption that $u^{0,0}$ does not depend on $v$.  It is then straightforward to check that a solution of the second equation is given by
\begin{equation}
r^{0,1}(t,z,v)=\frac{v\partial_zu^{0,0}(t,z)}{f(z)}.
\end{equation}
Finally, for any fixed $z$, let $\mu_z=\mathcal{N}(0,\sigma^2h(z)^2f(z))$ denote the invariant distribution of the Ornstein-Uhlenbeck process solving the SDE $\mathrm{d}V_z=-f(z)V_z\mathrm{d}t+\sigma\sqrt{2}f(z)h(z)\mathrm{d}W(t)$. The PDE satisfied by $u^{0,0}$ is obtained by taking the average of the last equation of the hierarchy, with respect to $\mathrm{d}\mu_z(v)$, and using the property $\int\overline{\A}_3\psi(v)\mathrm{d}\mu_z(v)=0$ for any smooth function $\psi$. Using that $\int v^2 \mathrm{d}\mu_z(v)=\sigma^2h(z)^2f(z)$, one obtains
\begin{equation}
\begin{aligned}
\partial_tu^{0,0}(t,z)&=\int \partial_tu^{0,0}(t,z)d\mu_z(v)\\
&=\int \A_1r^{0,1}(t,z,v)d\mu_z(v)\\
&=\sigma^2h(z)^2f(z)\partial_z\bigl(\frac{\partial_zu^{0,0}}{f(z)}\bigr)-\frac{g(z)}{f(z)}\partial_zu^{0,0}\\
&=-\Bigl(\frac{\sigma^2h(z)^2f'(z)}{f(z)}+\frac{g(z)}{f(z)}\Bigr)\partial_zu^{0,0}+\sigma^2h(z)^2\partial_{zz}^{2}u^{0,0}\\
&=\overline{\LL}^{0,0}u^{0,0}.
\end{aligned}
\end{equation}
The origin of the noise-induced drift term when $f$ is not constant appears clearly in the computation above. For completeness, for fixed $t$ and $z$, the function $r^{0,2}(t,z,\cdot)$ is constructed as solution of the Poisson equation
\begin{equation}
-\overline{\A}_3r^{0,2}(t,z,v)=\A_1r^{0,1}(t,z,v)-\int \A_1r^{0,1}(t,z,\cdot)\mathrm{d}\mu_z,
\end{equation}
which is solvable since the right-hand side is centered with respect to $\mu_z$.

The limiting generator $\overline{\LL}^{0,0}$ is associated with the SDE written in It\^o form
\begin{equation}
\mathrm{d}Z=-\frac{g(Z)}{f(Z)}\mathrm{d}t-\frac{\sigma^2h(Z)^2f'(Z)}{f(Z)}\mathrm{d}t+\sigma\sqrt{2}h(Z)\mathrm{d}W(t)
\end{equation}
where $\bigl(W(t)\bigr)_{t\ge 0}$ is a standard real-valued Wiener process.

The Stratonovich form of the SDE is written as
\begin{equation}
\mathrm{d}Z=-\frac{g(Z)}{f(Z)}\mathrm{d}t-\frac{\sigma^2 h(Z)(fh)'(Z)}{f(Z)}\mathrm{d}t+\sigma\sqrt{2}h(Z)\circ \mathrm{d}W(t).
\end{equation}

\subsection{Analysis in Regime 3}

In Regime 3, the parameters $\delta$ and $\epsilon$ go to $0$, with the constraint $\delta=\lambda\epsilon$, where $\lambda\in(0,\infty)$ is held fixed. In the sequel, we consider $\epsilon$ as the unique small parameter. Let $u_\lambda^\epsilon=u^{\epsilon,\lambda\epsilon}$.

Using the relation $\delta=\lambda\epsilon$, the infinitesimal generator $\LL^{\epsilon,\delta}$ given by~\eqref{eq:generator} is written as
\begin{equation}\label{eq:generator_regime3}
\LL_\lambda^\epsilon=\frac{1}{\epsilon}\B_1+\frac{1}{\epsilon^2}\B_{2,\lambda},
\end{equation}
where $\B_1=\A_1$ and $\B_{2,\lambda}=\frac{1}{\lambda}\A_2+\A_3+\frac{1}{\lambda^2}\A_4$.

One needs to construct an asymptotic expansion in terms of the small parameter $\epsilon$, of the form
\begin{equation}
u_\lambda^{\epsilon}(t,z,v,\zeta)=u_\lambda(t,z)+\epsilon r_{\lambda}^{1}(t,z,v,\zeta)+\epsilon^2r_\lambda^{2}(t,z,v,\zeta) + {\rm O}\left(\epsilon^{3} \right),
\end{equation}
where the zero-order term $u_\lambda$ does not depend on $v$ and $\zeta$ and describes the limiting process. Then, one needs to identify the limiting generator $\overline{\LL}_\lambda$ such that one has $\partial_tu_\lambda=\overline{\LL}_\lambda u_\lambda$. Inserting the asymptotic expansion in the backward Kolmogorov equation~\eqref{eq:Kolmogorov} and using the expression~\eqref{eq:generator_regime3} of the infinitesimal generator $\LL_\lambda^\epsilon$, one obtains the following hierarchy of equations when matching terms of size $\epsilon^{-2}$, $\epsilon^{-1}$ and $1$ respectively:
\begin{equation}
\begin{aligned}
\B_{2,\lambda}u_\lambda&=0 ,\\
\B_{2,\lambda}r_\lambda^1+\B_1u_\lambda&=0 ,\\
\B_{2,\lambda}r_\lambda^2+\B_1r_\lambda^1&=\partial_tu_\lambda.
\end{aligned}
\end{equation}
The first equation is consistent with the assumption that $u_\lambda$ does not depend on $v$ and $\zeta$.

The infinitesimal generator $\B_{2,\lambda}$ is associated with the two-dimensional SDE system for the components $v$ and $\zeta$, with frozen position component $z$:
\begin{equation}\label{eq:SDE_B2_lambda}
\begin{cases}
\mathrm{d}V_{\lambda,z}=-f(z)V_{\lambda,z}\mathrm{d}t+\frac{\sigma\sqrt{2}f(z)h(z)}{\lambda}\zeta_\lambda \mathrm{d}t\\
\mathrm{d}\zeta_\lambda=-\frac{\zeta_\lambda}{\lambda^2}\mathrm{d}t+\frac{1}{\lambda}\mathrm{d}\beta(t).
\end{cases}
\end{equation}
The process $(V_{\lambda,z},\zeta_\lambda)$ is a two-dimensional Ornstein-Uhlenbeck process, which converges when $t\to\infty$ to a centered Gaussian distribution $\mu_{\lambda,z}$ with covariance matrix characterized by
\begin{equation}
\begin{aligned}
\int \zeta^2 \mathrm{d}\mu_{\lambda,z}(v,\zeta)&=\frac12,\\
\int v\zeta \mathrm{d}\mu_{\lambda,z}( v,\zeta)&=\frac{\lambda \sigma f(z)h(z)}{\sqrt{2}(1+\lambda^2f(z))},\\
\int v^2 \mathrm{d}\mu_{\lambda,z}(v,\zeta)&=\frac{\sigma^2 h(z)^2f(z)}{1+\lambda^2 f(z)}.
\end{aligned}
\end{equation}
In fact, $\langle \zeta^2\rangle_{\lambda,z}=\int \zeta^2 \mathrm{d}\mu_z(v,\zeta)$, $\langle v\zeta\rangle_{\lambda,z}=\int v\zeta \mathrm{d}\mu_z( v,\zeta)$ and $\langle v^2\rangle_z=\int v^2 \mathrm{d}\mu_{\lambda,z}(v,\zeta)$ are obtained in the large time limit, and solve the system (derived for instance by Shapiro-Loginov procedure)
\begin{equation}
\begin{cases}
0=-\frac{\langle \zeta^2\rangle_{\lambda,z}}{\lambda^2}+\frac{1}{2\lambda^2},\\
0=-\bigl(\frac{1}{\lambda^2}+f(z)\bigr)\langle v\zeta\rangle_{\lambda,z}+\frac{\sigma\sqrt{2}f(z)h(z)}{\lambda}\langle \zeta^2\rangle_{\lambda,z},\\
0=-f(z)\langle v^2\rangle_{\lambda,z}+\frac{\sigma\sqrt{2}f(z)h(z)}{\lambda}\langle v\zeta\rangle_{\lambda,z}.
\end{cases}
\end{equation}

Define
\begin{equation}
r_\lambda^1(z,v,\zeta)=\frac{\partial_zu_\lambda}{f(z)}v+\lambda\sigma\sqrt{2}h(z)\partial_zu_\lambda\zeta,
\end{equation}
then one has $\B_{2,\lambda}r_\lambda^1+\B_1u_\lambda=0$. To identify the generator of the limiting SDE, it suffices to exploit the identity $\int \B_{2,\lambda}\psi(v,\zeta)\mathrm{d}\mu_{\lambda,z}(v,\zeta)=0$ for all smooth functions $\psi$, and to compute from the last equation of the hierarchy 
\begin{equation}
\begin{aligned}
\partial_tu_\lambda(t,z)&=\int \partial_t u_\lambda(t,z) \mathrm{d}\mu_{\lambda,z}(v,\zeta)\\
&=\int \B_1r_\lambda^1(t,z,v,\zeta)\mathrm{d}\mu_{\lambda,z}(v,\zeta)\\
&=-\frac{g(z)}{f(z)}\partial_zu_\lambda+\langle v^2\rangle_{\lambda,z}\partial_z\bigl(\frac{\partial_zu_\lambda}{f(z)}\bigr)+\lambda\sigma\sqrt{2}\langle v\zeta\rangle_{\lambda,z}\partial_z\bigl(h(z)\partial_zu_\lambda\bigr)\\
&=-\frac{g(z)}{f(z)}\partial_zu_\lambda\\
&~+\frac{\sigma^2 h(z)^2f(z)}{1+\lambda^2 f(z)}\partial_z\bigl(\frac{\partial_zu_\lambda}{f(z)}\bigr)+\frac{\lambda^2\sigma^2 h(z)f(z)}{1+\lambda^2 f(z)}\partial_z\bigl(h(z)\partial_zu_\lambda\bigr)\\
&=-\frac{g(z)}{f(z)}\partial_zu_\lambda+\sigma^2h(z)^2\partial_{zz}^2u_\lambda+\sigma^2h(z)h'(z)\partial_zu_\lambda\\
&~-\frac{\sigma^2(fh)'(z)h(z)}{f(z)(1+\lambda^2f(z))}\partial_zu_\lambda\\
&=\overline{\LL}_\lambda u_\lambda .
\end{aligned}
\end{equation}
The limiting generator $\overline{\LL}_\lambda$ is associated with the SDE written in It\^o form
\begin{equation}
\begin{aligned}
\mathrm{d}Z&=-\frac{g(Z)}{f(Z)}\mathrm{d}t+\sigma^2h(Z)h'(Z)\mathrm{d}t\\
&~-\frac{\sigma^2 h(Z)(fh)'(Z)}{(1+\lambda^2f(Z))f(Z)}\mathrm{d}t+\sigma\sqrt{2}h(Z)\mathrm{d}W(t),
\end{aligned}
\end{equation}
where $\bigl(W(t)\bigr)_{t\ge 0}$ is a standard real-valued Wiener process. One checks that the Stratonovich form of the SDE is
\begin{equation}
\mathrm{d}Z=-\frac{g(Z)}{f(Z)}\mathrm{d}t-\frac{\sigma^2 h(Z)(fh)'(Z)}{(1+\lambda^2f(Z))f(Z)}\mathrm{d}t+\sigma\sqrt{2}h(Z)\circ \mathrm{d}W(t).
\end{equation}

\end{appendix}

\bibliography{dustclouds}

\label{lastpage}
\end{document}

%% file: journaux.tex
%
%  These Macros are taken from the AAS TeX macro package version 4.0.
%  Include this file in your LaTeX source only if you are not using
%  the AAS TeX macro package and need to resolve the macro definitions
%  in the BibTeX entries returned by the ADS abstract service.
%
%  For more information on the AASTeX macro package, please see the URL
%	http://www.aas.org/publications/aastex.html
%  For more information about ADS abstract server, please see the URL
%	http://adswww.harvard.edu/ads_abstracts.html
%

% Abbreviations for journals.  The object here is to provide authors
% with convenient shorthands for the most "popular" (often-cited)
% journals; the author can use these markup tags without being concerned
% about the exact form of the journal abbreviation, or its formatting.
% It is up to the keeper of the macros to make sure the macros expand
% to the proper text.  If macro package writers agree to all use the
% same TeX command name, authors only have to remember one thing, and
% the style file will take care of editorial preferences.  This also
% applies when a single journal decides to revamp its abbreviating
% scheme, as happened with the ApJ (Abt 1991).

\def\jnl@style{\it}
%commente par Seb
\def\aaref@jnl#1{{\jnl@style#1}}
%ref remplace par aaref pour eviter conflit...

\def\aaref@jnl#1{{\jnl@style#1}}

\def\actaa{\aaref@jnl{Acta~Astronomica}}
\def\aj{\aaref@jnl{AJ}}                   % Astronomical Journal
\def\araa{\aaref@jnl{ARA\&A}}             % Annual Review of Astron and Astrophys
\def\apj{\aaref@jnl{ApJ}}                 % Astrophysical Journal
\def\apjl{\aaref@jnl{ApJ}}                % Astrophysical Journal, Letters
\def\apjs{\aaref@jnl{ApJS}}               % Astrophysical Journal, Supplement
\def\ao{\aaref@jnl{Appl.~Opt.}}           % Applied Optics
\def\apss{\aaref@jnl{Ap\&SS}}             % Astrophysics and Space Science
\def\aap{\aaref@jnl{A\&A}}                % Astronomy and Astrophysics
\def\aapr{\aaref@jnl{A\&A~Rev.}}          % Astronomy and Astrophysics Reviews
\def\aaps{\aaref@jnl{A\&AS}}              % Astronomy and Astrophysics, Supplement
\def\azh{\aaref@jnl{AZh}}                 % Astronomicheskii Zhurnal
\def\baas{\aaref@jnl{BAAS}}               % Bulletin of the AAS
\def\icarus{\aaref@jnl{icarus}} 
\def\jrasc{\aaref@jnl{JRASC}}             % Journal of the RAS of Canada
\def\memras{\aaref@jnl{MmRAS}}            % Memoirs of the RAS
\def\mnras{\aaref@jnl{MNRAS}}             % Monthly Notices of the RAS
\def\pra{\aaref@jnl{Phys.~Rev.~A}}        % Physical Review A: General Physics
\def\prb{\aaref@jnl{Phys.~Rev.~B}}        % Physical Review B: Solid State
\def\prc{\aaref@jnl{Phys.~Rev.~C}}        % Physical Review C
\def\prd{\aaref@jnl{Phys.~Rev.~D}}        % Physical Review D
\def\pre{\aaref@jnl{Phys.~Rev.~E}}        % Physical Review E
\def\prl{\aaref@jnl{Phys.~Rev.~Lett.}}    % Physical Review Letters
\def\pasp{\aaref@jnl{PASP}}               % Publications of the ASP
\def\pasj{\aaref@jnl{PASJ}}               % Publications of the ASJ
\def\qjras{\aaref@jnl{QJRAS}}             % Quarterly Journal of the RAS
\def\skytel{\aaref@jnl{S\&T}}             % Sky and Telescope
\def\solphys{\aaref@jnl{Sol.~Phys.}}      % Solar Physics
\def\sovast{\aaref@jnl{Soviet~Ast.}}      % Soviet Astronomy
\def\ssr{\aaref@jnl{Space~Sci.~Rev.}}     % Space Science Reviews
\def\zap{\aaref@jnl{ZAp}}                 % Zeitschrift fuer Astrophysik
\def\nat{\aaref@jnl{Nature}}              % Nature
\def\iaucirc{\aaref@jnl{IAU~Circ.}}       % IAU Cirulars
\def\aplett{\aaref@jnl{Astrophys.~Lett.}} % Astrophysics Letters
\def\apspr{\aaref@jnl{Astrophys.~Space~Phys.~Res.}}
                % Astrophysics Space Physics Research
\def\bain{\aaref@jnl{Bull.~Astron.~Inst.~Netherlands}} 
                % Bulletin Astronomical Institute of the Netherlands
\def\fcp{\aaref@jnl{Fund.~Cosmic~Phys.}}  % Fundamental Cosmic Physics
\def\gca{\aaref@jnl{Geochim.~Cosmochim.~Acta}}   % Geochimica Cosmochimica Acta
\def\grl{\aaref@jnl{Geophys.~Res.~Lett.}} % Geophysics Research Letters
\def\jcp{\aaref@jnl{J.~Chem.~Phys.}}      % Journal of Chemical Physics
\def\jgr{\aaref@jnl{J.~Geophys.~Res.}}    % Journal of Geophysics Research
\def\jqsrt{\aaref@jnl{J.~Quant.~Spec.~Radiat.~Transf.}}
                % Journal of Quantitiative Spectroscopy and Radiative Transfer
\def\memsai{\aaref@jnl{Mem.~Soc.~Astron.~Italiana}}
                % Mem. Societa Astronomica Italiana
\def\nphysa{\aaref@jnl{Nucl.~Phys.~A}}   % Nuclear Physics A
\def\physrep{\aaref@jnl{Phys.~Rep.}}   % Physics Reports
\def\physscr{\aaref@jnl{Phys.~Scr}}   % Physica Scripta
\def\planss{\aaref@jnl{Planet.~Space~Sci.}}   % Planetary Space Science
\def\procspie{\aaref@jnl{Proc.~SPIE}}   % Proceedings of the SPIE

\let\astap=\aap
\let\apjlett=\apjl
\let\apjsupp=\apjs
\let\applopt=\ao